\documentclass[aps,preprint,preprintnumbers,nofootinbib,showpacs,prd]{revtex4}
\pdfoutput=1
\usepackage{graphicx,color,amsmath,amsfonts,multirow}
\usepackage{epsfig}
\usepackage{rotating}
\usepackage{array}
\usepackage{graphicx}
\usepackage{multirow}
\usepackage{amssymb}
\usepackage{latexsym}
\usepackage{ulem}

\newcommand{\ba}{\begin{array}}
\newcommand{\be}{\begin{eqnarray}}
\newcommand{\ee}{\end{eqnarray}}
\newcommand{\bit}     {\begin{itemize}}
\newcommand{\eit}     {\end{itemize}}
\newcommand{\ben}     {\begin{enumerate}}
\newcommand{\een}     {\end{enumerate}}
\newcommand{\bad}{\begin{array}{ccc}}
\newcommand{\ea}{\end{array}}

\newcommand{\bbt}{\bibitem}
\newcommand{\lsim}{\mathrel{\mathop{\kern 0pt \rlap
  {\raise.2ex\hbox{$<$}}}
  \lower.9ex\hbox{\kern-.190em $\sim$}}}
\newcommand{\gsim}{\mathrel{\mathop{\kern 0pt \rlap
  {\raise.2ex\hbox{$>$}}}
  \lower.9ex\hbox{\kern-.190em $\sim$}}}

\newcommand{\sg}      {\sigma}

\newcommand{\tth}{{ ttH}}
\newcommand{\tR}{\hat{\mu}}

\newcommand{\fig}{\begin{figure}}
\newcommand{\ef}{\end{figure}}

\newcommand{\vbf}{ {\rm VBF}}

\newcommand{\gev}{{\;{\rm GeV}}}

\newcommand{\gtwo}{ {(g-2)_\mu} }

\preprint{PITT-PACC 1514, ~P15059}

\begin{document}
\title{Muon $g-2$ in the Aligned Two Higgs Doublet Model
}

\author{Tao Han$^{a,c,d}$, Sin Kyu Kang$^b$, Joshua Sayre$^a$}

\affiliation{
$^a$Department of Physics and Astronomy, University of Pittsburgh, Pittsburgh, PA 15260, USA\\
$^b$School of Liberal Arts, Seoul National University of Science and Technology, Seoul 139-743, Korea \\
$^c$ Physics Department, Collaborative Innovation Center of Quantum Matter, Tsinghua University, Beijing 100084, China\\
$^d$Korea Institute for Advanced Study, Seoul 130-722, Korea 
}

\date{\today}
\begin{abstract}
We study the Two-Higgs-Doublet Model with the aligned Yukawa sector (A2HDM) in light of the observed excess measured in the muon anomalous magnetic moment. 
We take into account the existing theoretical and experimental constraints with up-to-date values and demonstrate that a phenomenologically interesting region of parameter space exists. 
With a detailed parameter scan, we show a much larger region of viable parameter space in this model beyond the limiting case Type X 2HDM as obtained before.
It features the existence of light scalar states with masses $3\gev \lsim m_H^{} \lsim 50\gev$, or $\ 10\gev\lsim m_A^{} \lsim 130\gev$, with enhanced couplings to tau leptons. The charged Higgs boson is typically heavier, with $200\gev \lsim m^{}_{H^+} \lsim 630 \gev.$
The surviving parameter space is forced into the CP-conserving limit by EDM constraints. Some Standard Model observables may be significantly modified, including a possible new decay mode of the SM-like Higgs boson to four taus. We comment on  future measurements and direct searches for those effects at the LHC as tests of the model.

\end{abstract}

\maketitle
\section{Introduction}
\label{sec:introduction}

After the discovery of a Higgs boson with mass around 125 GeV~\cite{ATLAS-Higgs,CMS-Higgs,ATLAS-7,CMS-7}, the ATLAS and CMS collaborations have continued to accumulate data and measured 
the detailed properties of this new particle. Within the attainable accuracy, they appear
consistent with the elementary Higgs boson of the Standard Model (SM).
This milestone discovery strongly encourages the exploration for new physics, in particular the search for a richer structure of the Higgs sector beyond the SM. Although no clear indication exists yet for new physics beyond the SM, 
this search will be actively pursued at the LHC and elsewhere.

The apparent discrepancy with a $3-4$ sigma deviation between 
the theoretical \cite{Aoyama:2012wk,Czarnecki:2002nt,g-2-review1,g-2-review2}
and measured \cite{g-2-Collab} values of the muon anomalous magnetic moment $(g-2)_\mu$ is a long-standing puzzle which may point to new physics \cite{Hagiwara:2011,Davier:2010,Benayoun:2012,Blum:2013xva,Recent}. 
The possibility that the observed excess in $\gtwo$ can be explained by an extended Higgs sector has been raised by several authors \cite{Dedes:2001nx,Cao:2009as,Broggio:2014mna,Wang:2014sda,Abe:2015oca,Hektor:2015zba,Review}.
%
Extended Higgs sectors arise in a variety of new physics scenarios, including 
supersymmetry, Grand Unified Theories, dark matter models, flavor models, and others. A feature common to many models, and one of the simplest extensions, is the addition of a second Higgs doublet. We thus will focus on a general formulation of 
the two-Higgs-doublet-model (2HDM)~\cite{Haber:1978jt,2hdm1,2hdm2,2hdm3,Gunion:1989we,Branco:2011iw,Dev:2015bta} in this work in the hope to reach some general conclusions. 

2HDMs have been extensively studied in the past, most often in the context of global flavor symmetries and with the assumption of no new sources of CP violation. 
In this paper we focus on an interesting broader formulation, the Aligned Two Higgs Doublet Model (A2HDM) \cite{flavour-2hdm}. We find that the A2HDM can account for the experimental value of $\gtwo$ \cite{g-2-Collab}.
However, it can only account for such an excess in a very restricted range of the otherwise-allowed parameter space. In particular, mixing between the new states and the SM Higgs boson must be very small, one of the new neutral states must be quite light, the new states will couple strongly to taus, and CP-violating effects must be negligible. 

The paper is organized as follows: We describe the A2HDM model for our study in Sec.~\ref{sec:review}, and then present the current experimental bounds on the model parameters in Sec.~\ref{sec:constraints}.
Accounting for these constraints, we perform a detailed multiple-dimensional parameter scan in light of the $\gtwo$ excess, and show our main results in Sec.~\ref{sec:results}. We discuss the physical implication of these results on SM observables in Sec.~\ref{sec:SM} and comment on direct searches for the new Higgs states at the LHC in Sec.~\ref{sec:search}.
In  Sec.~\ref{sec:conclude}, we draw our conclusions. An appendix is provided to summarize our $\gtwo$ calculation in the A2HDM model.
 
\section{Brief review of A2HDM}
\label{sec:review}

For two SU(2)$_L$ Higgs doublets with  hyper-charge one, we can redefine 
a theoretical basis by rotating the doublets according to a global $U(2)$ transformation \cite{basis-independent}. We choose to work in the Higgs basis in which only one doublet, $H_1$, contains the SM electroweak symmetry breaking vacuum expectation value (vev) \cite{higgs-basis}:
\begin{eqnarray}
\langle H^{0}_{1} \rangle =\frac{v}{\sqrt{2}}, ~~~\langle H^{0}_{2} \rangle=0.
\end{eqnarray}
The scalar potential can then be expressed using Higgs basis fields as follows:
\begin{eqnarray}
V&=&Y_1 H^{\dagger}_{1}H_1+Y_2 H^{\dagger}_{2}H_2+[Y_3 H^{\dagger}_{1}H_2+h.c] \nonumber\\
&+&\frac{1}{2}Z_1 (H^{\dagger}_{1}H_1)^{2}+\frac{1}{2}Z_2 (H^{\dagger}_{2}H_2)^{2}
+Z_3 (H^{\dagger}_{1}H_1) (H^{\dagger}_{2}H_2)+Z_4 (H^{\dagger}_{1}H_2) (H^{\dagger}_{2}H_1)
\nonumber \\
&+&\left\{\frac{1}{2}Z_5 (H^{\dagger}_{1}H_2)^{2}+[Z_6 (H^{\dagger}_{1}H_1)+Z_7 (H^{\dagger}_{2}H_2)] H^{\dagger}_{1}H_2+h.c \right\},
\end{eqnarray}
where $Y_{1}, Y_{2}$ and $Z_{1,2,3,4}$ are real-valued and invariant in this basis, whereas
$Y_{3}$ and $Z_{5,6,7}$ are complex ``pseudo-invariants". The pseudo-invariants can be re-phased by the $U(1)$ transformation 
$H_2 \to e^{i \chi}H_2$.
The SM-like minimization condition for the scalar potential requires
\begin{eqnarray}
Y_{1}=-\frac{1}{2}Z_1 v^2, ~~~Y_{3}=-\frac{1}{2}Z_6 v^2.
\end{eqnarray}

The neutral Higgs boson mass-eigenstates can be determined by diagonalizing a $3\times 3$ squared-mass  matrix  given by
\begin{eqnarray}
M^2 = v^2\left(\begin{array}{ccc}
  Z_1 & \mbox{Re}(Z_6 e^{-i\theta_{23}}) & -\mbox{Im}(Z_6 e^{-i\theta_{23}}) \\
\mbox{Re}(Z_6 e^{-i\theta_{23}}) & A^2/v^2 +\mbox{Re}(Z_5 e^{-2i\theta_{23}}) &
-\frac{1}{2}\mbox{Im}(Z_5 e^{-2i\theta_{23}}) \\
-\mbox{Im}(Z_6 e^{-i\theta_{23}}) & -\frac{1}{2}\mbox{Im}(Z_5 e^{-2i\theta_{23}}) &
A^2/v^2 \end{array} \right),
\end{eqnarray} 
where $A^2 \equiv Y_2+\frac{1}{2}[Z_3+Z_4-\mbox{Re}(Z_5e^{-2i\theta_{23}})] v^2.$
The diagonalizing matrix $R$, 
\begin{eqnarray}
R M^2 R^{T} = M^2_{D}=\mbox{diag}[m^2_1, m_2^2, m_3^2],
\end{eqnarray} can be written as 
\begin{eqnarray}
R= \left(\begin{array}{ccc}
 c_{12} c_{13} &-s_{12} & -c_{12}s_{13} \\
s_{12}c_{13} & c_{12} & - s_{12} s_{13} \\
s_{13} & 0 & c_{13}  \end{array} \right),
\end{eqnarray} 
where $c_{ij}=\cos\theta_{ij}$ and $s_{ij}=\sin\theta_{ij}$.
In general $\theta_{23}$ appears in the diagonalizing matrix $R^{'}$ for the matrix $M^{'}$ in some particular 
pseudo-invariant basis.
 Under the rephasing $H_2 \rightarrow e^{i\chi} H_2$, however,
 $\theta_{23}\rightarrow \theta_{23}-\chi$ \cite{haber}. Thus $R$ and $M$ as written are invariant and we may use the remaining freedom in our choice of basis to eliminate $\theta_{23}$, which has no physical significance. 
Here, we choose a convention where $-\frac{1}{2}\pi \leq \theta_{12}, \theta_{13}<\frac{1}{2}\pi$.
We will identify one of the mass eigenstates as the observed SM-like Higgs near 125 GeV.
The charged Higgs mass is given by
\begin{eqnarray}
m^2_{H^{\pm}}=Y_2+\frac{1}{2}Z_3v^2.
\end{eqnarray}

The Higgs couplings to the fermions can be written as
\be
-L_Y=&&\bar{U}_L \left( \kappa^u H_1^{0 \dagger} + \rho^u H_2^{0 \dagger}\right) U_R 
-\bar{D}_L K^\dagger \left(\kappa^u H_1^- + \rho^u H_2^-\right)U_R \nonumber \\
+&& \bar{D}_L \left( \kappa^{d\dagger} H_1^{0 \dagger} + \rho^{d\dagger} H_2^{0 \dagger}\right) D_R 
-\bar{U}_L K \left(\kappa^{d\dagger} H_1^+ + \rho^{d^\dagger} H_2^+\right)D_R \nonumber \\
+&& \bar{L}_L \left( \kappa^{l\dagger} H_1^{0 \dagger} + \rho^{l\dagger} H_2^{0 \dagger}\right) L_R 
-\bar{\nu}_L  \left(\kappa^{l\dagger} H_1^+ + \rho^{l^\dagger} H_2^+\right)L_R . 
\ee
where $K$ is the CKM matrix and $\kappa^i =\frac{\sqrt{2}}{v}M_i$ is proportional to the diagonal SM masses 
\be
M_U=\mbox{diag}[m_u, m_c, m_t], ~~ M_D=\mbox{diag}[m_d, m_s, m_b],~~ M_L=\mbox{diag}[m_e, m_\mu, m_\tau].
\ee
The A2HDM is defined by the relation \cite{flavour-2hdm}
\be
\label{eq:a2hdm}
\rho_i = A_i^* \kappa_i
\ee
where $A_i$ is a potentially complex number. Thus the Yukawa couplings of the two Higgs doublets are aligned\footnote{Note that this alignment is different from the ``alignment limit" discussed in certain literature which corresponds with small mixing between $H_1$ and 
$H_2$ \cite{ATwoHDM}.} in that they are proportional to each other and can be simultaneously diagonalized, which ensures that flavor changing neutral currents (FCNCs) do not arise at tree level. 
The A2HDM is thus an example of Minimal Flavor Violation \cite{MFV}. 
FCNC effects will develop from loop diagrams involving the charged Higgs, which always depend on powers of the CKM matrix. The often studied discrete $Z_2$ 2HDMs can be understood as special cases of this model which correspond with the relations among the $A_i$ parameters summarized in Table \ref{tab:aim}. 
In these cases, the ratio of Yukawa couplings  allows us to rotate to a basis where the symmetry is manifest and only one doublet couples to each fermion type (up, down and lepton, respectively). In such cases, $|A_u|$ is equivalent to $\cot \beta \equiv  v_1/v_2$, the ratio of vevs in that basis.

\begin{table}
\centering
\begin{tabular}{|c|c|c|}
\hline
&$A_u=A_d$ & $A_u = -(A_d^*)^{-1}$ \\ \hline
$A_l=A_d$ & Type I & Type II \\ \hline
$A_l=-(A_d^*)^{-1}$ & Type X & Type Y \\ \hline
\end{tabular}
\caption{Parameter relations corresponding to models with a discrete symmetry.}
\label{tab:aim}
\end{table}

In the 2HDM, CP is potentially broken in the scalar potential and vacuum as well as in the neutral Higgs Yukawa 
interactions. CP is preserved if the following terms are real \cite{haber}
\begin{eqnarray}
Z_5 (Z_6^*)^2,~~~Z_6Z_7^*,~~~Z_5 (\rho^i)^2, ~~~Z_6 \rho^i, ~~~Z_7 \rho^i,
\end{eqnarray}
where $i=u,d$ and $l$. In our chosen basis this is equivalent to setting $Z_5,Z_6,Z_7,A_u,A_d,A_l$ to be real and it implies that $s_{13}=0$.

Details of the general CP-violating A2HDM and calculations relevant to our scan can be found 
in a forthcoming publication Ref.~\cite{cppaper}.


\section{Constraints on A2HDM}
\label{sec:constraints}

There exist a number of relevant theoretical and experimental constraints on the A2HDM. For theoretical considerations, we require that the quartic coupling parameters $Z_i$ satisfy the partial-wave unitarity bounds, which are taken from the Appendix of \cite{haber}. We also require that the SM-like electroweak symmetry breaking vev be at a local minimum, and that the potential be positive at large field values. For convenience, we set the parameter $Z_2=4\pi$, corresponding to the maximum value allowed by partial-wave unitarity. $Z_2$ does not enter any of our phenomenological considerations so we choose a value that contributes maximally to the stability of the potential.

On the experimental side, we take the following results into account: 

\subsection{Precision EW Data}

We directly impose the LEP bound on the charged Higgs mass 
~\cite{Agashe:2014kda}
\be
m_{H^\pm} \geq 80\gev.
\ee
For other searches applicable to exotic Higgs, we make use of the program HiggsBounds \cite{Bechtle:2013wla}, which checks for exclusion of potential  signals at the $95\%$ level incorporating a large number of searches from the LEP, Tevatron, LHC and other experiments. We also check for compatibility with the Peskin-Takeuchi $S,~T$ and $U$ parameters \cite{Peskin:1990zt} at the $Z$ pole. Experimentally, the allowed values for these parameters are \cite{Agashe:2014kda}
\begin{eqnarray}
 S = -0.03 \pm 0.10, \quad
T = 0.01 \pm 0.12, \quad
U= 0.05 \pm 0.10. 
\end{eqnarray}

A further constraint of interest for the large $|A_l|$ values we allow comes from measurements of the $Z \tau \overline{\tau}$ coupling measured at the $Z$ pole. This is characterized by effective couplings $g_\tau^L$ and $g_\tau^R$ which have been determined to high precision. Experimentally they are found to be \cite{ALEPH:2005ab}:
\be
g_\tau^L &= -0.26930 \pm 0.00058, \quad
g_\tau^R &= 0.23208 \pm 0.00062.
\ee
SM best fit predictions based on precision data yield \cite{Agashe:2014kda}
\be
g_\tau^L(SM) &= -0.26919 \pm 0.0002, \quad
g_\tau^R(SM) &= 0.23274 \pm 0.0002.
\ee
Our calculation for the effects of the new Higgs states is based on Ref.~\cite{Logan:1999if}.

\subsection{$B$ Meson Rare Decays}

Typically, some of the strongest experimental constraints on the 2HDM come from flavor physics \cite{flavor, flavor-typeII,flavor-general}. Although highly suppressed in the SM, FCNCs are experimentally observed and provide stringent constraints on new physics. 
The A2HDM guarantees that tree-level FCNCs vanish, but they will appear at the loop level. We make use of the NLO prediction of $BR(B\rightarrow X_s \gamma)$ in the 2HDM available from the SusyBSG code 
provided by the authors of Ref.~\cite{bsg-NLO}. The numerical estimation of $BR(B\rightarrow X_s \gamma)$ has been performed up to NNLO for the SM \cite{bsg-NNLO}.
The most updated SM prediction is 
\begin{eqnarray}
BR(b\rightarrow s \gamma)|_{\rm SM}= (3.15 \pm 0.23)\times 10^{-4}.
\end{eqnarray}
The most recent average of experimental data on $b\rightarrow s \gamma$ rate gives \cite{Agashe:2014kda}
\begin{eqnarray}
BR(b\rightarrow s \gamma)|_{\rm exp}= (3.43 \pm 0.22)\times 10^{-4}.
\end{eqnarray}
NNLO corrections for the 2HDM are discussed in Ref.~\cite{2hdm-NNLO}. 

A second rare decay of interest is $B_s \to \mu^{+}\mu^{-}$. This decay is of high theoretical significance and may receive contributions from both charged and neutral Higgs bosons in the A2HDM via box and penguin diagrams. Details of the calculation can be found in Ref.~\cite{bsmu-2hdm}. The SM prediction works out to be
\be
BR(B_s \to \mu^+\mu^-)_{\rm SM} = (3.67 \pm 0.25)\times 10^{-9}.
\ee
Measurements of this decay have been made at both LHCb \cite{bsmu-lhcb} and CMS \cite{bsmu-cms} and found to be consistent with the SM expectation. The combined result can 
be expressed in a ratio for comparison with the Standard Model:
 \be
R_{B_s \to \mu\mu} \equiv \frac{BR(B_s \to \mu^+\mu^-)_{exp} }{BR(B_s \to \mu^+\mu^-)_{\rm SM} }= 0.79 \pm 0.20.
\ee

\subsection{Lepton Universality}
Many additional flavor phenomena can potentially be affected by new physics in the A2HDM. A large selection of 
relevant processes is analyzed in Ref.~\cite{flavour-2hdm} in the A2HDM under the assumption that the charged scalar $H^+$ provides the dominant effects. These include leptonic tau decays, leptonic decays of heavy mesons such as $B \to \tau \nu$ and $D_{(s)} \to \mu(\tau) \nu$, measurements of $Z \to b \overline{b}$, $B^0-\overline{B^0}$ mixing and $K^0-\overline{K^0}$ mixing.
These can be translated into constraints at $95\%$ confidence on the charged Higgs mass and fermion couplings as follows:
\be
\label{eq:flavor}
&&|A_l| < 0.4 M_{H^+};\ \ 
|A_u| < 0.56+2.65 \cdot 10^{-3} M_{H^+} -1.05 \cdot 10^{-6} M_{H^+}^2 +6.15 \cdot 10^{-10} M_{H^+}^3 ; ~~~\quad \\ \nonumber
&&|A_l^* A_u| < 0.005 M_{H^+}^2; \quad
 -0.036 M_{H^+}^2 < A_l^* A_d < 0.008 M_{H^+}^2 
 \ \ \ {\rm (for~real ~Yukawa ~couplings).}
\ee

The most relevant of these limits for our purposes is the first, which comes from tests of lepton flavor universality in tau decays. 
 In particular, it derives from the ratio of tau decay widths $\overline{\Gamma} (\tau \to \mu \nu \overline{\nu}) / \overline{\Gamma} (\tau \to e \nu \overline{\nu})$, where $\overline{\Gamma}$ indicates the partial width normalized to its SM value. This ratio is equivalent to the ratio of fitted effective coupling parameters $(g_\mu/g_e)^2$ given by the HFAG collaboration \cite{hfag}. In Ref.~\cite{Abe:2015oca}, the authors make use of the additional fitted value  $(g_\tau/g_e)^2$ which can be translated into the ratio $\overline{\Gamma} (\tau \to \mu \nu \overline{\nu}) / \overline{\Gamma} (\mu \to e \nu \overline{\nu})$.  The most recent HFAG fits to purely leptonic decays are
\be
\label{eq:hfag}
\frac{g_\mu}{g_e} = 0.0018 \pm 0.0014, \qquad \frac{g_\tau}{g_e} = 0.0029 \pm 0.0015, \qquad \frac{g_\tau}{g_\mu} = 0.0011 \pm 0.0015.
\ee

 Neglecting the electron Yukawa coupling, only $\tau \to \mu \nu$ is altered by  the tree-level charged Higgs graph.Within the measured bounds of the Michel parameters which characterize the decay distribution, the only allowable effects from the tree-level charged Higgs will tend to decrease $\Gamma (\tau \to \mu \nu \overline{\nu})$. Thus the bound depends strongly on the fact that the current fit to $(g_\tau/g_e)^2$ is almost a $2\sigma$ excess over the Standard Model while the charged Higgs gives small negative corrections.

In light of this, Ref.~\cite{Abe:2015oca} includes the leading one-loop effects from the new Higgs states which tend to decrease the coupling of $W \tau \nu$. The one-loop effect applies equally to both tau decays. Hence, the ratio $g_\mu/g_e$ is only affected by the tree-level Higgs graph, while $g_\tau/g_e$ should include both tree-level and leading one-loop effects. They exclude parameters which exceed the $95\%$ confidence limits on the combined fit, accounting for correlations in the data. We adopt their calculation of the one-loop effects, which is consistent with Ref.~\cite{Krawczyk:2004na}. 

In principle we may work with the ratio $g_\tau/g_\mu$ instead of $g_\tau/g_e$ since only two of the ratios in Eq.~(\ref{eq:hfag}) are independent. The ratio $g_\tau/g_\mu$ is equivalent to $\overline{\Gamma} (\tau \to e \nu \overline{\nu}) / \overline{\Gamma} (\mu \to e \nu \overline{\nu})$ and recieves a negative correction from the one-loop graphs but not from 
the tree-level charged Higgs. Hence, the strength of the bound comes from the fact that both  $g_\tau/g_\mu$ and $g_\mu/g_e$ are high in the purely leptonic fits. However, HFAG also provides fits to  $g_\tau/g_\mu$ from 
$\overline{\Gamma} (\tau \to h\nu) / \overline{\Gamma} (h \to \mu \overline{\nu})$ where $h = K,~\pi$, which are found to be low compared to the SM expectation. These ratios should also be affected by the one loop corrections but not the tree level term since we neglect the Higgs coupling to light quarks. The combined fit is then reported as \cite{hfag}
\be
(\frac{g_\tau}{g_\mu})_{\tau + \pi + K} = 0.0001 \pm 0.0014.
\ee
We use this number to constrain the one loop corrections at $95\%$ confidence and $g_\mu/g_e$ to constrain the tree-level charged Higgs correction. This gives a weaker exclusion than the purely leptonic fits.

\subsection{Heavy Quarkonium Decay}

As will be seen, our results allow for rather light new neutral scalars which may have a non-negligible coupling to heavy quarks. 
For scalars with masses below $10$ GeV we consider bounds from measurements of rare $\Upsilon$ decays. In particular, the branching fraction for $\Upsilon (1S) \to \gamma \mu^+\mu^-$  with a narrow resonance in $\mu^+\mu^-$ and similarly for $\Upsilon (3S) \to \gamma \tau^+ \tau^-$  can constrain the coupling of new light scalars to $b$ quarks \cite{McKeen:2008gd}. Based on Refs.~\cite{Lees:2012iw,Aubert:2009cka} we impose the following limits:
\be
&& BR(\Upsilon (1S) \to \gamma \phi ) \times BR(\phi \to \mu^+ \mu^-) < 10^{-6}\times (m_{\phi}/\text{GeV}), \\
&& BR(\Upsilon (3S) \to \gamma \phi ) \times BR(\phi \to \tau^+ \tau^-) < 3 \times 10^{-5}.
\ee

\begin{table}
\begin{tabular}{|c|l|l|}\hline
decay channel & ~~~~~ATLAS & ~~~~~CMS \\\hline
$\gamma \gamma $ 
& $\tR^{ggF} = 1.32\pm 0.38$ \cite{Higgs:diphoton:2014:ATLAS}
& $\tR^{ggF+\tth} = 1.07^{+0.37}_{-0.31}$ \cite{Khachatryan:2014jba,Higgs:diphoton:2014:CMS} \\
& $\tR^{\vbf} = 0.8\pm0.7$ \cite{Higgs:diphoton:2014:ATLAS}  
& $\tR^{\vbf} = 1.24^{+0.63}_{-0.58}$ \cite{Khachatryan:2014jba,Higgs:diphoton:2014:CMS}  \\
\hline
$WW$
& $\tR^{ggF} = 1.02^{+0.29}_{-0.26}$ \cite{ATL-2013-13}
& $\tR^{ggF} = 0.87^{+0.21}_{-0.21}$ \cite{Khachatryan:2014jba,1312.1129} \\
& $\tR^\vbf = 1.27^{+0.53}_{-0.45}$ \cite{ATL-2013-030}
& $\tR^\vbf = 0.66^{+0.5}_{-0.46}$ \cite{Khachatryan:2014jba,1312.1129} \\
\hline
$ZZ$
& $\tR^{ggF+\tth} = 1.7^{+0.5}_{-0.4}$ \cite{Aad:2014eva} 
& $\tR^{ggF+\tth} = 0.88^{+0.46}_{-0.36}$ \cite{Khachatryan:2014jba,1312.5353} \\
& $\tR^{\vbf+VH} = 0.3^{+1.6}_{-0.9}$ \cite{Aad:2014eva}
& $\tR^{\vbf+VH} = 1.75^{+2.2}_{-2.1}$ \cite{Khachatryan:2014jba,1312.5353} \\
\hline
$\tau \tau$ 
& $\tR^{ggF} = 2.0^{+1.47}_{-1.17}$ \cite{Aad:2015vsa} 
& $\tR^{ggF} = 0.52\pm 0.4$ \cite{Khachatryan:2014jba,1401.5041} \\
& $\tR^{\vbf+VH} = 1.24^{+0.59}_{-0.54}$ \cite{Aad:2015vsa}
& $\tR^\vbf = 1.21\pm0.5$ \cite{Khachatryan:2014jba,1401.5041} \\
\hline
$b\bar{b}$
& $\tR^{VH} = 0.52^{+0.4}_{-0.4}$ \cite{Aad:2014xzb} 
& $\tR^{VH} = 0.85\pm0.5$ \cite{Khachatryan:2014jba,3687} \\
\hline
\end{tabular}
\caption{Summary of the LHC Higgs signals at 7 and 8 TeV.}
\label{table:LHCdata}
\end{table}

\subsection{Bounds from SM-like Higgs Searches at the LHC}

Experimental measurements of the Higgs boson properties at the LHC have been typically characterized by the so-called signal strength modifiers defined by $\hat\mu^i = \sigma^i/\sg_{\rm SM}$.
The recent measured values of the signal strength modifiers for each channel under consideration  are summarized in Table \ref{table:LHCdata}, where the superscripts of the signal strength modifiers denote the production channels of the Higgs boson.

We perform a global $\chi^2$ fit of the model predictions
to the observed Higgs signal strengths $\tR^i$. We assume each channel listed can be treated as an independent measurement, giving us 18 degrees of freedom from the LHC measurements as listed in  Table \ref{table:LHCdata}.
Note that some analyses in the literature give separate fits to associated ($VH$) and vector-boson fusion (VBF) production. In a given model, they are not independent. In our treatment we use the combined fit, taking into account the fact that these two processes are determined by the scalar mixing parameter $\theta_{12}$ for both $W$s and $Z$s. 
Similarly, some analyses have performed a fit to associated top quark ($t\overline{t}H$) production, but at present these have large error bars and we choose to use the combined fit with gluon-gluon fusion ($ggF$) when appropriate since the latter is dominated by the top quark loop. We require 
a predicted $\chi^2$ value consistent with experiment at the $95\%$ level.

In the framework of the A2HDM, the SM-like Higgs boson could decay to two lighter states when kinematically accessible, which then subsequently decay to four taus. An estimate of exclusion limits for such decays based on recast 3-lepton searches was made in Ref.~\cite{Curtin:2013fra}. They found that $BR(h \to AA \to 4\tau) < 25 \%$ for $m_A > 30$ GeV, with weaker bounds for $m_A < 30$ GeV. 
Recently, CMS has presented a search for this decay mode with boosted kinematics for the $A$s which excludes  $BR(h \to AA \to 4\tau) \gtrsim 25 \%$ in the range $8<m_A < 15$ \cite{cms4tau}. The light states may also decay to muons which makes the process sensitive to bounds from $h \to AA \to 2\tau 2\mu$ searches. It turns out that these limits have become the strongest for lower masses with $BR(h \to AA) \lesssim 10 \%$ for $m_A \sim 10$ GeV \cite{atlas2tau2mu}. If the new particles are too light to decay to taus then we expect them to decay largely to muons, which is bounded by experiment to be less than $\sim 1~ \text{fb}$ \cite{cms4mu}. 
We include these limits as a constraint on our results.

\subsection{Bounds from non-SM Searches at the LHC}
Additional scalar particles introduce the possibility of many exotic signals as the LHC and other experiments. In general 
we employ the HiggsBounds code to check for consistency of our model with experiment. However, we have directly incorporated limits from two recent updates which are particularly pertinent to the A2HDM with large $\gtwo$. One is the direct search for 
$gg \to H/A \to \tau \overline{\tau}$ \cite{CMStauupdate}. The second is the search for $pp \to (H/A) \to (A/H)Z \to 2\tau 2 l$ \cite{HZtauupdate}.

\begin{figure}[!tb]
\begin{centering}
\includegraphics[width=9cm]{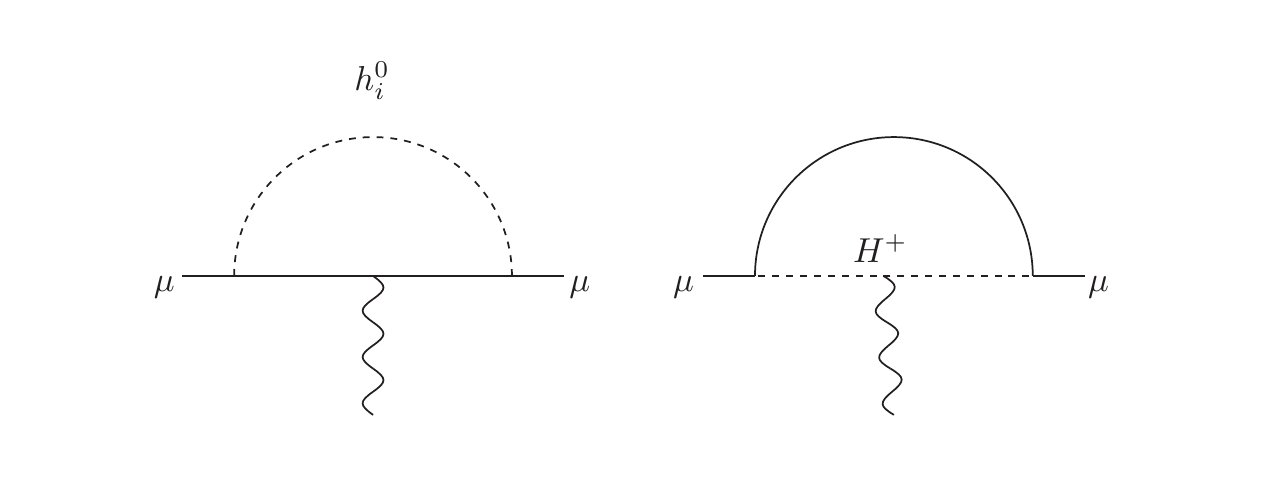}\ 
\includegraphics[width=10cm]{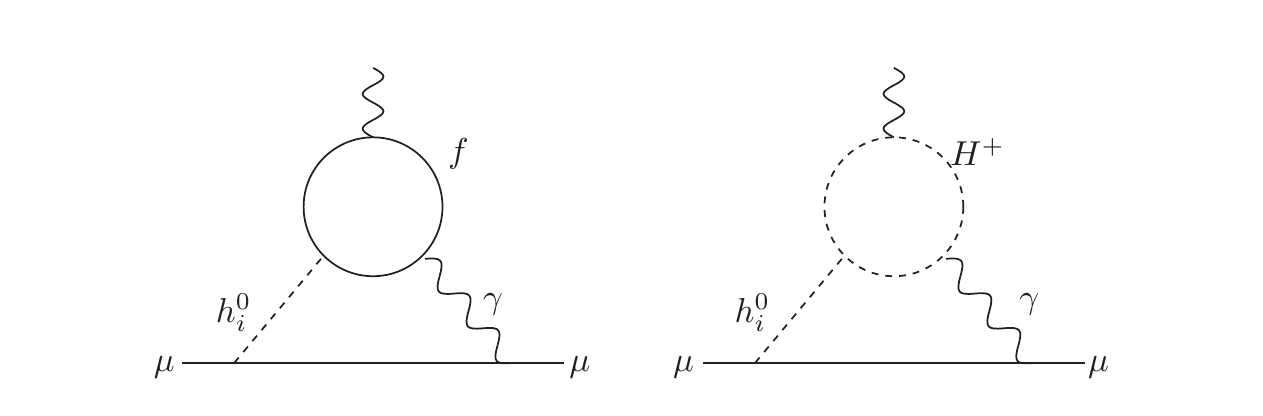}
\caption{Representative Feynman diagrams in 2HDM contributing to $\gtwo$.}
\label{fig:diagrams}
\end{centering}
\end{figure}

\subsection{Measurements of $\gtwo$}

The SM value of the muon anomalous magnetic moment has been calculated up to five-loop 
contributions in QED \cite{Aoyama:2012wk} and two-loop in weak interactions \cite{Czarnecki:2002nt,g-2-review1,g-2-review2}. Among the diagrams, the hadronic contributions are particularly challenging to reliably estimate \cite{Blum:2013xva}.
The current SM prediction for  $a_{\mu}^{\rm SM} \equiv (g-2)_\mu/2$ results in a $(3-4) \sigma$ deviation from the experimental result from Brookhaven E821 \cite{g-2-Collab,Blum:2013xva}
\begin{eqnarray}
\Delta a_\mu \equiv (a^{\rm EXP}_{\mu} - a^{\rm SM}_{\mu}) = (261 \pm 78) \times 10^{-11}.
\end{eqnarray}  
In the SM, the Higgs contribution to $a_{\mu}$ is suppressed by a factor of $m^2_{\mu}/m^2_h$
compared to the electroweak contributions \cite{g-2-review1,g-2-review2}.
However, the Higgs sector contributions to $a_{\mu}$ can be considerably enhanced in a 2HDM. The significance of  the $a_\mu$ constraint for 2HDMs was emphasized in Refs.~\cite{g-22hdm,Dedes:2001nx,Cheung:2003pw}. 
Representative Feynman diagrams in the A2HDM contributing to $\gtwo$ are depicted in Fig.~\ref{fig:diagrams}. New Higgs bosons may contribute to $\gtwo$ at leading order at the one-loop level. 
However, for a Higgs boson mass larger than $\sim 10$ GeV, dominant Higgs contributions to  $(g-2)_\mu$ come from the two-loop Barr-Zee diagram with a fermion in the loop \cite{barr-zee}.
It is also known that in the CP-conserving 2HDM of Type X, a light pseudo-scalar, together with a large $\tan\beta$ value can  explain the measured $\Delta a_\mu$ via such diagrams \cite{Dedes:2001nx,Cao:2009as,Broggio:2014mna,Wang:2014sda,Abe:2015oca}.
Recently, the importance of additional contributions arising from charged Higgs bosons in the A2HDM has been emphasized in Ref.~\cite{Ilisie:2015tra}. We include these in our predictions.


\section {Results of the Numerical Scan}
\label{sec:results}

We perform a scan over the free parameters of the Higgs potential and Yukawa couplings. 
As mentioned above, without constraining our results $Z_2$ is fixed at $4\pi$.
We randomly generate points with a flat distribution over the range for each parameter allowed by unitarity and consistent with positive eigenvalues for the masses, specifically\footnote{Note these conditions are necessary but not sufficient and we make 
additional checks to ensure that a generated point satisfies our requirements.}, 
\begin{eqnarray}
&& 0< |A_u| <1.2,\quad  0<|A_d| < 50,\quad  0<|A_l| < 120 ,  \\
&& 0 < Z_1 < 4\pi,\  
-\sqrt{Z_1Z_2} < Z_3 < 8\pi,\  \frac{80^2}{v^2} < \frac{Y_2}{v^2} +\frac{Z_3}{2} < 4\pi + \frac{80^2}{v^2}, \\
&& Max[-8\pi,-2\frac{Y_2}{v^2}, \frac{125^2 -2Y_2}{v^2}-Z_1 ]< Z_3+Z_4 < 8 \pi,\\
&& 0 < |Z_5| < Min[2 \pi, \frac{2 Y_2}{v^2} +Z_3+Z_4],\  0 < |Z_7| < 2\pi.~~
\nonumber
\end{eqnarray}
Without loss of generality we then choose $|Z_6|$ so as to guarantee that one mass sits at $125$ GeV.
We record generated points which pass all the constraints discussed above within $2\sigma$ of the experimental values, including the requirement that the $\chi^2$ fit to LHC Higgs data is consistent within $95\%$ bounds.
We generate billions of points, most of which are discarded due to the stringent experimental constraints. We interpret our solutions only as representative since there may still be corners of parameter space not being fully sampled.

 \subsection{$\gtwo$ in the CP-conserving A2HDM}

In the CP-conserving model there are two CP-even ($h$ and $H$) and one CP-odd ($A$) mass eigenstates. We will label the SM-like state at $125$ GeV as $h$.
Points in the model parameter space which can facilitate a large positive $\Delta a_\mu$ correction are relatively rare.
This can be understood by considering the possible source of such corrections. Except for very light Higgs, the dominant contributions to $g-2$ in the 2HDM come from Barr-Zee diagrams with a fermion running in the loop \cite{barr-zee}. Large enhancements of $\Delta a_{\mu}$ thus require a relatively light Higgs with enhanced couplings to the muon or to the fermion in the loop. This can be achieved with large values of $A_u,\ A_d$ or $A_l$. However, the top-quark Yukawa coupling cannot be much larger than the SM-like value of unity without becoming non-perturbative and enhancement of the the coupling would lead to large rates of $b \to s \gamma$, $B_s \to \mu^+\mu^-$ and other flavor effects, as well as excessive production rates of the new Higgs bosons  via gluon fusion at the LHC. Similar problems would also   arise from the $b$-quark contributions if 
$|A_d| \gtrsim 40$.
Generally then, one must rely on an enhanced coupling to leptons with large $A_l$, which increases the contribution from all loops due to the muon coupling and doubly enhances the graph with a tau loop. 
This implies that light new Higgs states will typically decay to taus and any mixing of the 125 GeV Higgs with the scalar from $H_2$ will lead to significant deviations from an SM-like coupling to  taus, which is strongly constrained by current LHC results. Hence, $\gtwo$ forces us into the large $A_l$ region with small mixing angle $\theta_{12}$, which in turn requires that $Z_1 v^2$ is close to $(125\ \text{GeV})^2$ and $|Z_6|$ is small. A light Higgs with large $A_l$ also favors small $A_d$ so as not to violate various flavor bounds.

Note that the effect of scalars versus pseudo-scalars on $\gtwo$ is not as simple in the A2HDM as in the more restricted models which have been considered in the literature \cite{Cao:2009as,Broggio:2014mna,Wang:2014sda,Abe:2015oca}. At the one-loop level it is generically true that the pseudo-scalar gives a negative contribution to $\gtwo$ while scalars give a positive contribution. However, at two loops the sign of the contribution depends on the relative sign of the neutral Higgs coupling to the muon and to the fermion in the loop. Both Type-II and Type-X 2HDMs have the potential to positively enhance $\Delta a_\mu$ via two-loop diagrams.\footnote{Type-I and Type-Y models can also enhance $\gtwo$ in principle, but this requires large top-quark couplings which conflict with other phenomenological constraints.} In both models, however, the top quark loop cannot play the dominant role. This is because the pseudo-scalar $A$ couples to the muon  like $\tan \beta$ while it couples to the top 
like $\cot \beta$, which means that the contribution of such diagrams cannot be enhanced by the choice of $\tan \beta$.  In the small mixing limit the scalar $H$ has similar couplings. Thus only the bottom and tau loops can give a large enhancement in the Type-II model and only the taus in the Type-X model. This also fixes the sign of the potentially large contributions, so that the pseudo-scalar term is positive and the scalar $H$ term is negative.
In contrast, for the more general A2HDM model the top loop can  play a crucial role such that either scalars or pseudo-scalars can generate large contributions with a positive sign. Moreover, the contributions from charged Higgs, which have often been neglected, can have either sign.

\begin{figure}[!tb]
\begin{centering}
\includegraphics[width=8.1cm]{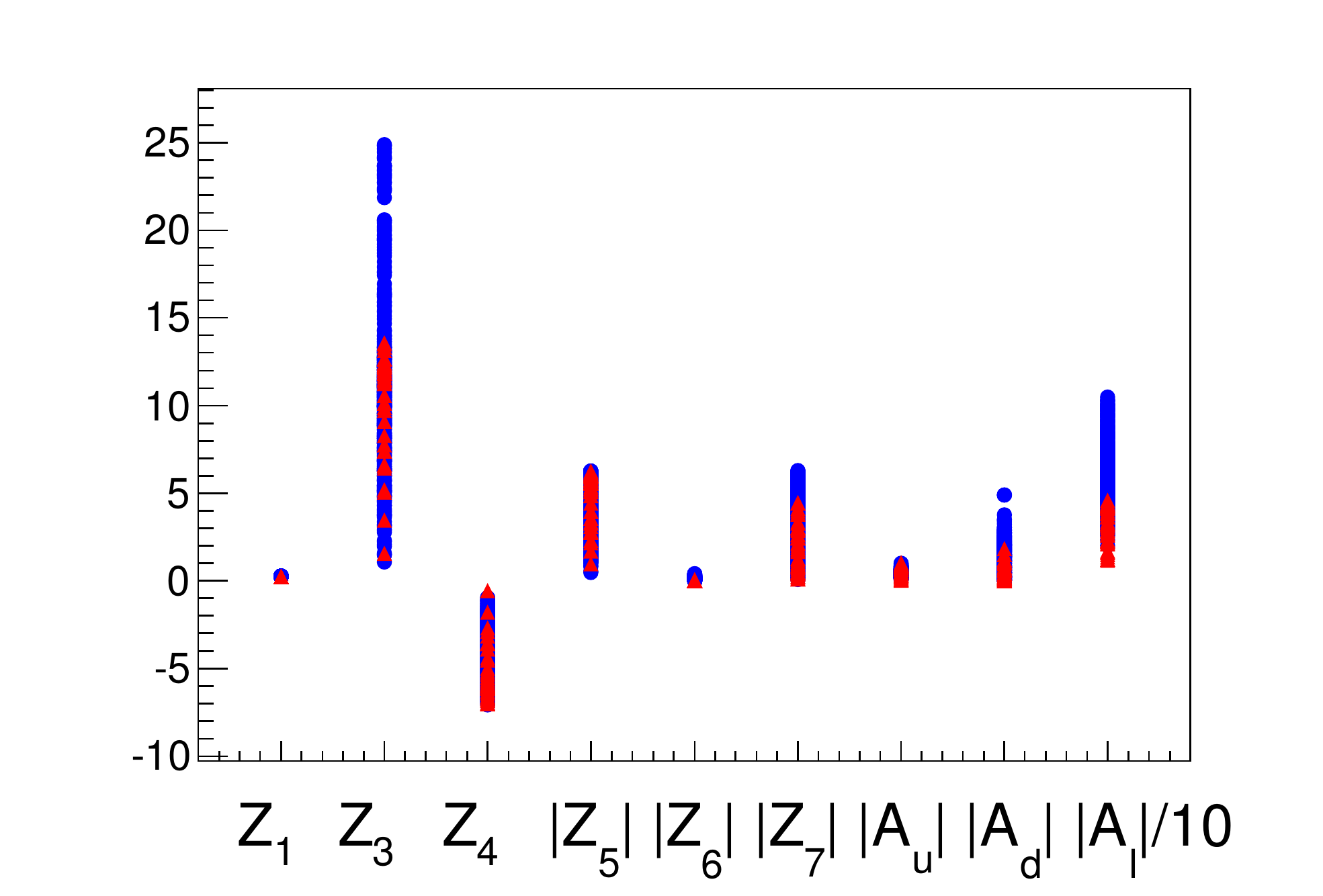}
\includegraphics[width=8.1cm]{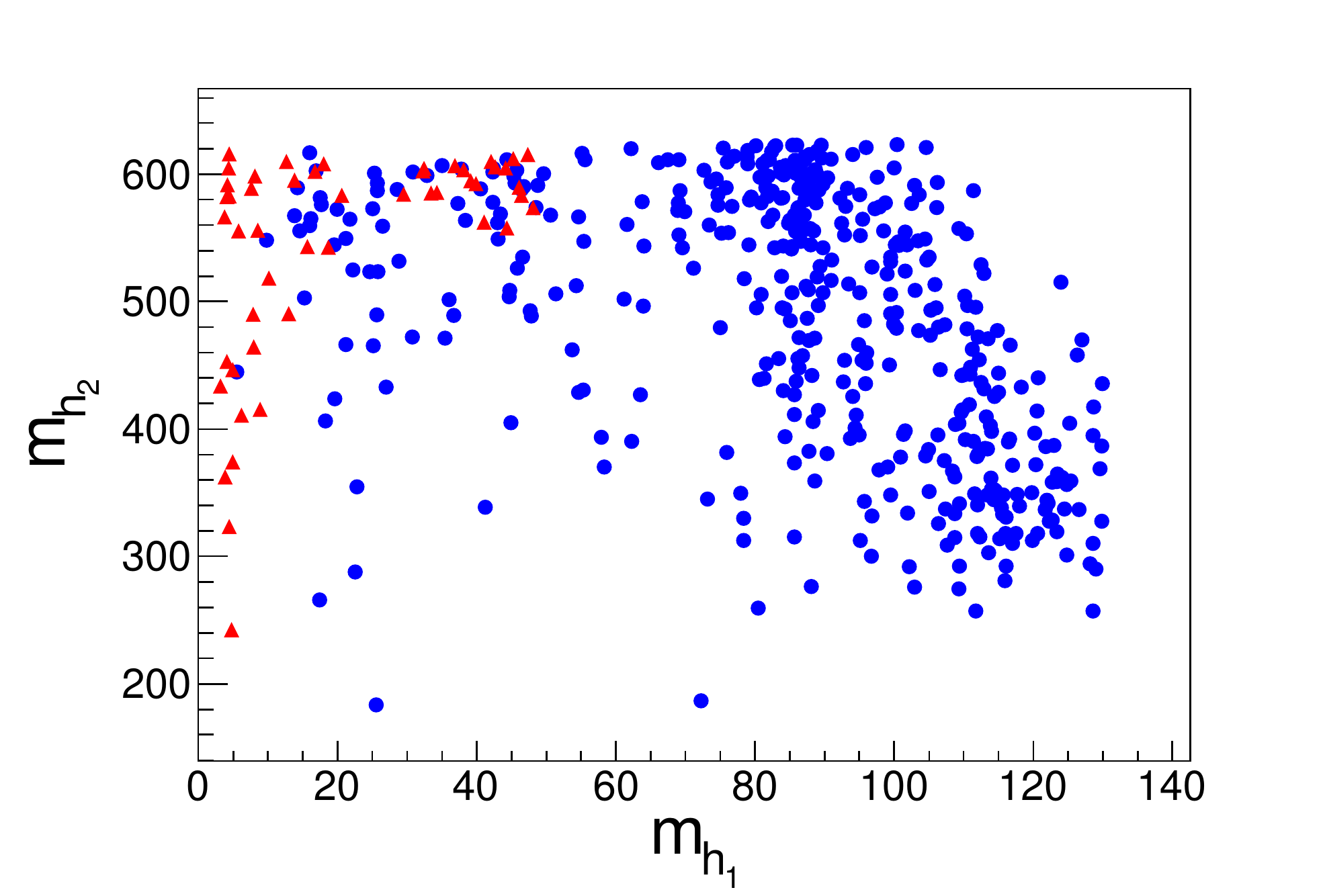}
\includegraphics[width=8.1cm]{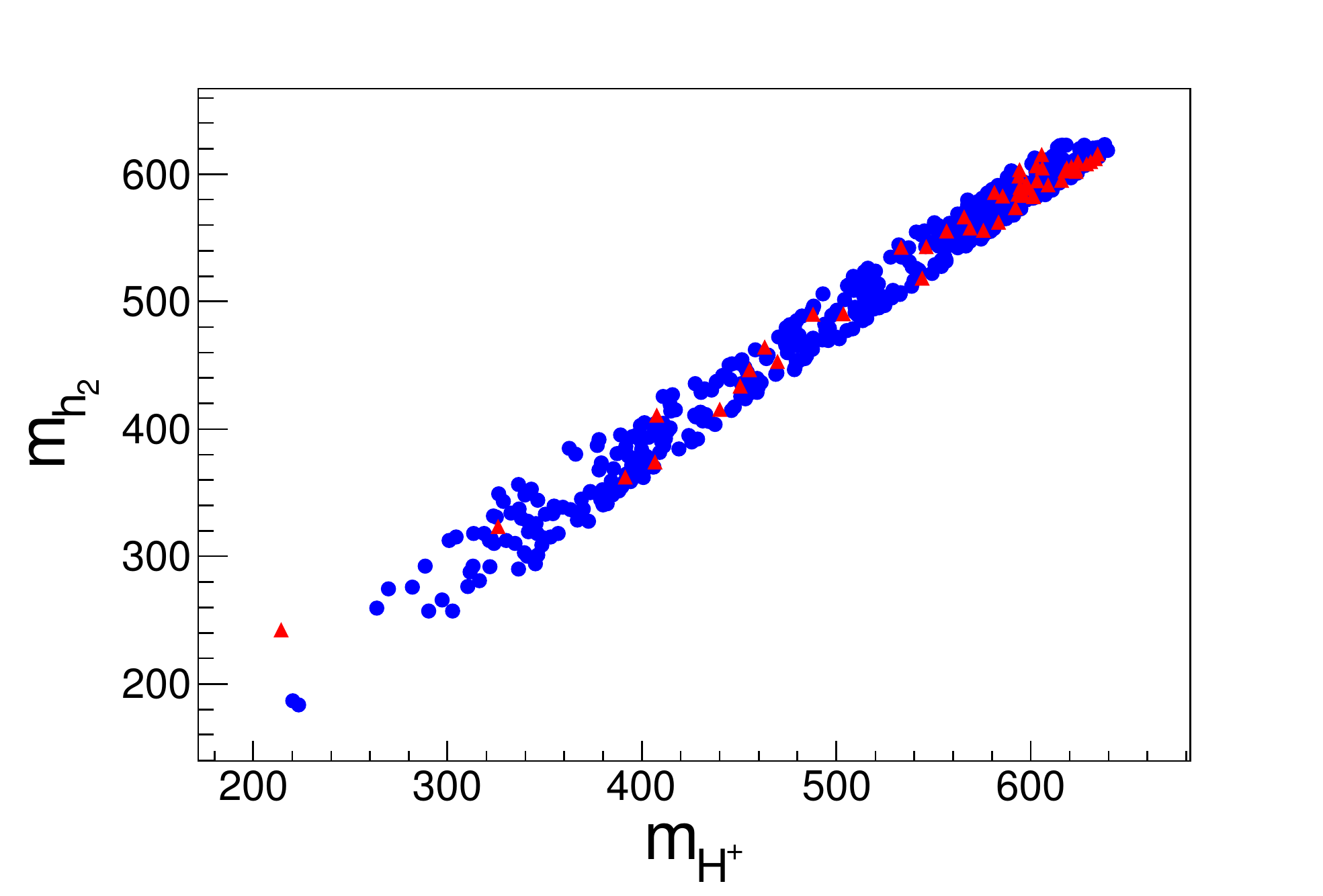}
\includegraphics[width=8.1cm]{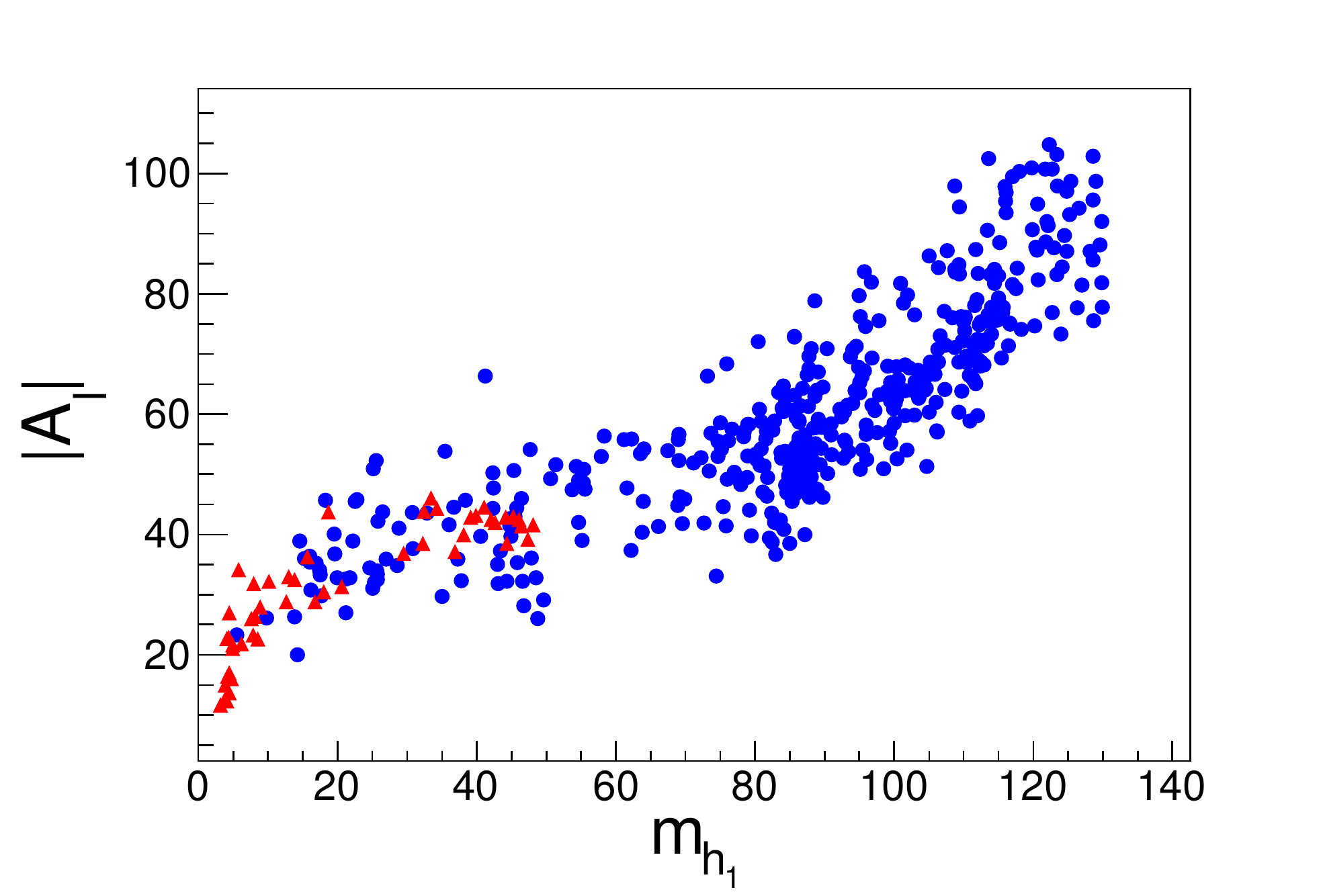}
\includegraphics[width=8.1cm]{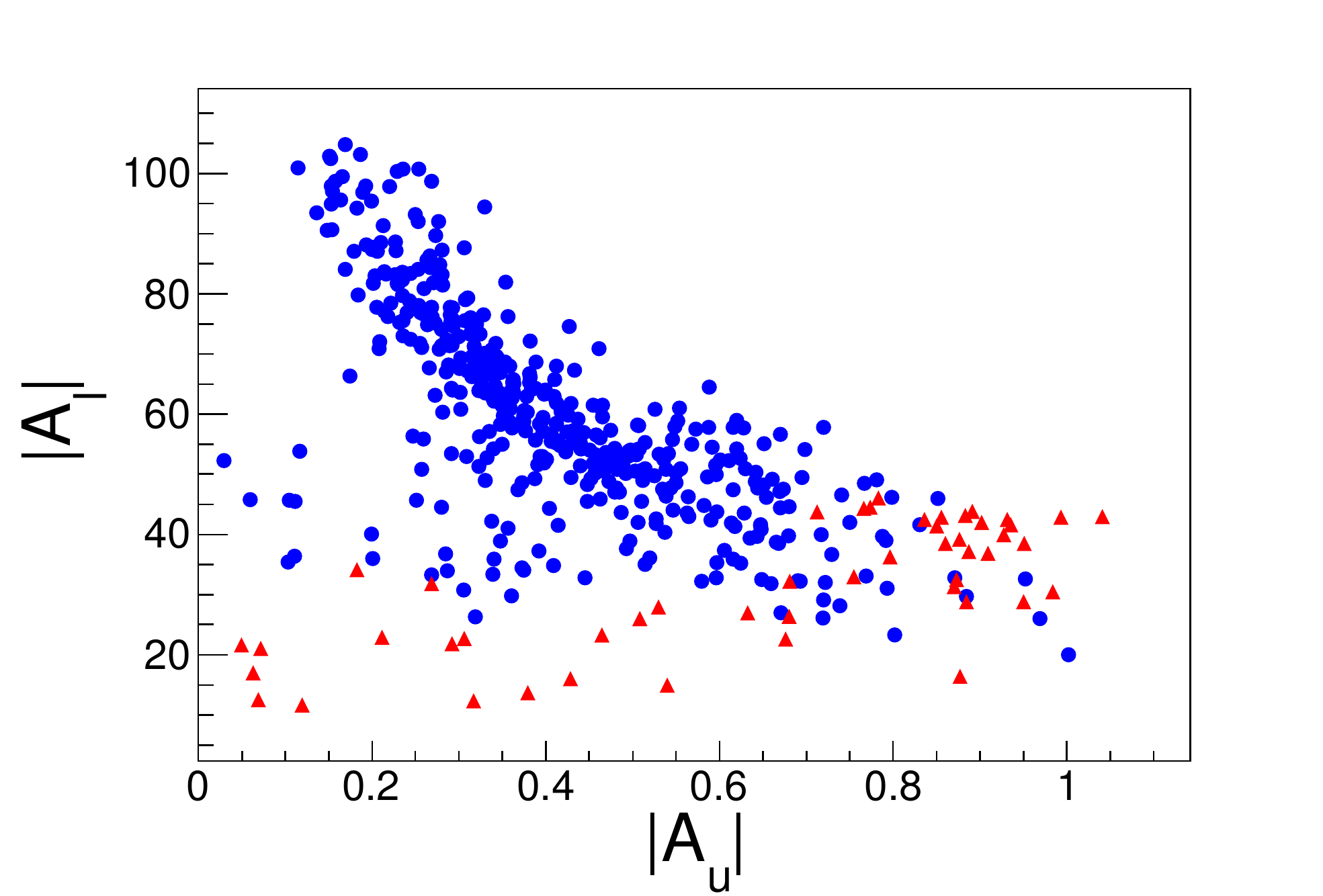}
\includegraphics[width=8.1cm]{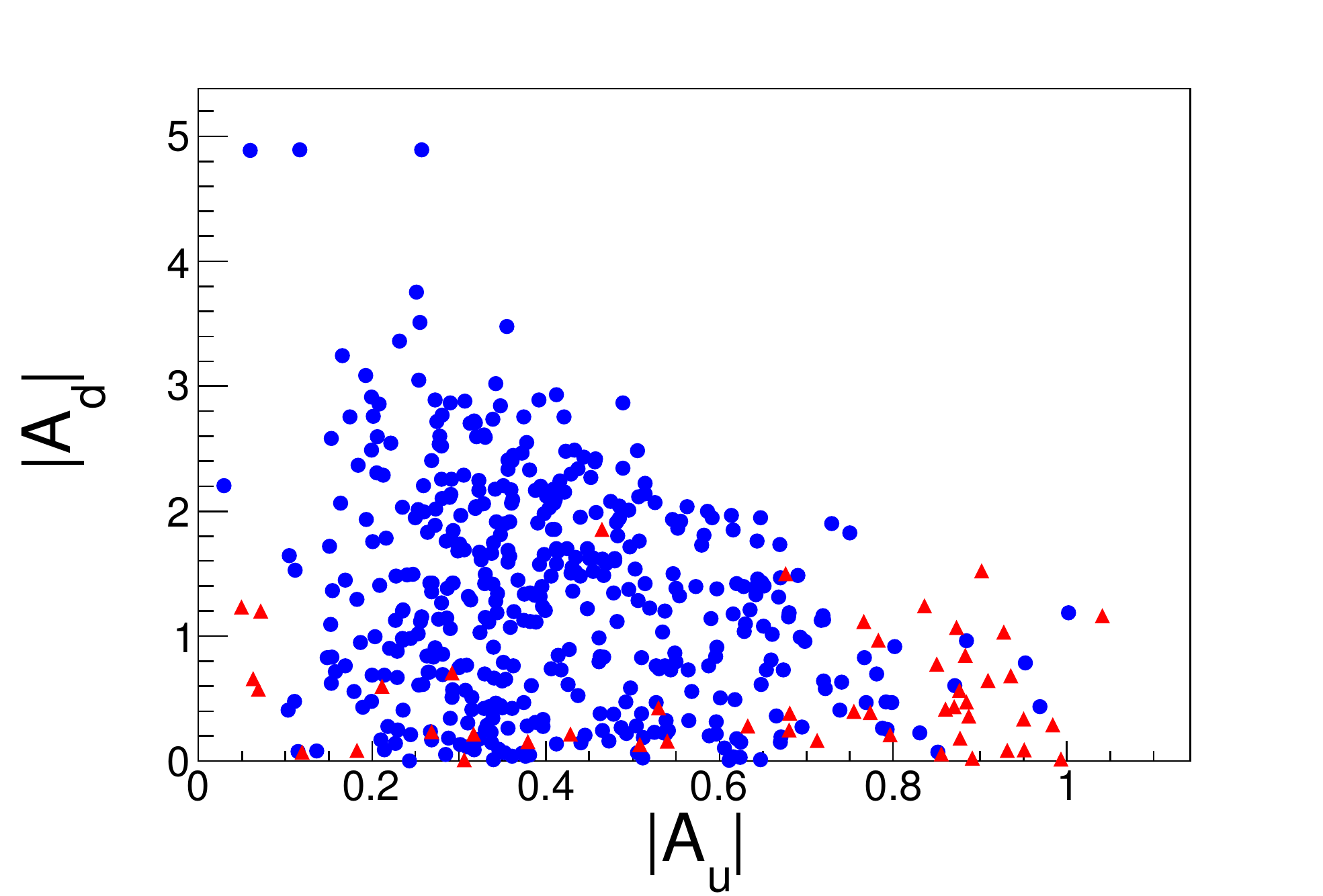}
\caption{Upper left: From left to right, allowed ranges of $Z_1$, $Z_3$, $Z_4$, $Z_5$, $Z_6$, $Z_7$, $|A_u|$, $|A_d|$, and $|A_l|$. 
Upper right:  Mass of the lighter new neutral state $h_1$ versus the heavier new neutral state $h_2$. 
Here and henceforth, blue circles  indicate a light pseudo-scalar and red triangles a light scalar. 
Middle left: Mass of the heavy new state $h_2$ versus the charged Higgs $H^\pm$. Middle right: Comparison of $m_{h_1}$ with $|A_l|$. Lower left: $|A_u|$ versus $|A_l|$. Lower right: $|A_u|$ versus $|A_d|$. All masses are given in GeV.}
\label{fig:zrange}
\end{centering}
\end{figure}

In the upper left panel of Fig.~\ref{fig:zrange} we show the ranges for input parameters which pass our constraints. As discussed, $Z_1$ is largely fixed, varying within about $15\%$ of the observed Higgs mass value $m_h^2/v^2 = 0.26$, and $Z_6$ is also small, $ \lesssim 0.5$, corresponding to the low-mixing limit. $Z_3$ is positive and can be as large as the unitarity bound at $8\pi$.
$Z_4$ is negative and $|Z_5|$ is approximately of the same magnitude. Given the small mixing, this follows from the requirement that one non-SM-like neutral state be approximately degenerate with the charge Higgs mass and the other new neutral state is significantly lighter. These two conditions allow the bounds on the $T$-parameter and the large $\gtwo$ to be satisfied. The viable range of $A_d$ is also much smaller than in the general case, where it can be as large as $\sim 50$, although it can still be mildly enhanced compared to the SM-like Yukawa coupling, up to a factor of $5$ approximately. 

The masses of the new neutral states which are determined from the values of $Z_i$  
are shown in the upper right panel of Fig.~\ref{fig:zrange}. We find it useful to distinguish the new neutral states by their mass ordering and adopt the notation $h_1$ for the lighter new neutral state and $h_2$ for the heavier. The SM-like Higgs at $125$ GeV will remain $h$ without a subscript. Here and henceforth, points with $m_H < m_A$ are shown as red triangles and with $m_A < m_H$ as blue circles. It is clear that there are three qualitative regions of interest:\footnote{The relative number of points in the different regions does not directly represent the relative volume of parameter space which they occupy.} 
\begin{eqnarray}
\label{eq:regions}
&&{\rm Low}\ m_A^{}:\ 10\gev \lsim m_A^{} \lsim 130 \gev, \quad 200\gev \lsim m_H^{} \sim m^{}_{H^+} \lsim 630 \gev ;  \\ 
&& {\rm Low}\ m_H^{}\text{(1-Loop):}\  3\gev  \lsim m_H^{} \lsim 20 \gev, \quad 200\gev \lsim m_A^{}\sim m^{}_{H^+} \lsim 630 \gev;~~~ \nonumber \\ 
&& {\rm Low}\ m_H^{}\text{(2-Loop):}\  30\gev  \lsim m_H^{} \lsim 50 \gev, \quad 550\gev \lsim m_A^{}\sim m^{}_{H^+} \lsim 630 \gev. \nonumber
\end{eqnarray}

\noindent
$\bullet$ 
In the first region, which occupies most of the allowed parameter space, the pseudo-scalar is quite light, less than $130$ GeV ranging down to about $10$ GeV. In this low $m_A$ region the new scalar can range from above $200$ GeV up to $630$ GeV and the charged Higgs has a similar mass. The  correlation of the charged mass with $m_{h_2}$  is shown in the middle left panel of Fig.~\ref{fig:zrange}. 
In this region, the dominant source of new $\gtwo$ contributions are 2-loop Barr-Zee graphs with tops or taus in the loop as in the lower left graph of Fig.~\ref{fig:diagrams}. 

\noindent
$\bullet$ 
A second region can be seen where the scalar has a very low mass, $m_H^{} = 3-20$ GeV and $A$ is heavy,  $m_A \sim 200-630$ GeV. Here, the charged Higgs is approximately degenerate with $A$. The leading contributions to $\gtwo$ arise at one-loop level from the top two graphs in Fig.~\ref{fig:diagrams}.

\noindent
$\bullet$ 
A small third region, with $\gtwo$ dominantly from the 2-loop Barr-Zee diagrams, can be distinguished when $m_A \gtrsim 550$ GeV where $H$ is somewhat heavier, in the range $30 < m_H < 50$ GeV.

In all regions, the upper bound on the mass of the heavier Higgs is determined 
by the allowed mass splitting between $H$ and $A$. This splitting is largely controlled by $|Z_5|$, which is constrained by unitarity to be less than $2\pi$. 

This pattern can be understood by considering the plot of $A_l$ versus $m_{h_1}$, shown in the middle right panel of Fig.~\ref{fig:zrange}.
The light pseudo-scalar region can account for a large $\gtwo$ value with $|A_l|$ as low as $\sim 25$. Most solutions involve a pseudo-scalar with mass around $100$ GeV, although they can be substantially lighter with  a corresponding decrease on the viable range of $|A_l|$. 
In general the large value of $|A_l|$ means that the lighter new (pseudo) scalar decays primarily to taus.
When 2-loop Barr-Zee diagrams with top loops dominate, either $H$ or $A$ can give a positive contribution to $\gtwo$. However, they are constrained by the $gg \to \phi \to \tau^+ \tau^-$ search. This is the primary constraint which disallows a heavier $A$. The sharp feature at $m_{h_1} = 90$ GeV arises from the fact that published results on this search begin at that mass. At high values of $|A_l|$ this bound is slightly weaker than at low values since the top coupling through $|A_u|$ can be reduced while still accounting for the large $\Delta a_\mu$. At the upper end of our allowed range for $|A_l|$ the pseudo-scalar 
mass can approach $130$ GeV with a light charged Higgs helping to enhance $\gtwo$.
Two light neutral particles which decay to taus and couple to quarks strongly enough to account for the magnetic moment are largely excluded. Thus when $A$ is light, the scalar $H$ must usually be significantly heavier, such that it can decay instead to $AA$, to $AZ$, or to top quarks. This implies that typically the heavier neutral state is above $200$ GeV. This bound also follows from the approximate degeneracy between the heavier neutral state and the charged Higgs. The latter is preferably heavier than the top quark to avoid constraints from top decays. 

If $|A_l|$ is large enough to generate $\Delta a_\mu$ through the tau loop alone, then in principle the quark couplings can be arbitrarily small, making the tau search from direct production irrelevant. This can be taken as the relevant limit for the Type X model. In this case, however, without the sizeable contribution of top loops to $\gtwo$, the pseudo-scalar must be lighter or $|A_l|$ even larger than the typical values shown in our plots. This parameter space is highly constrained by the measurements of the $Z\tau\bar{\tau}$ coupling and lepton universality. 
When tau loops dominate the scalar terms give a negative contribution to $\Delta a_\mu$ which must be suppressed relative to the pseudo-scalar, so a large mass splitting is still required.

The second region, with very light scalars and intermediate mass pseudo-scalars, corresponds with relatively mild enhancement of the lepton coupling, i.e. $|A_l|$ between $5$ and $35$. These scalars are light enough, $3-20$ GeV, that their positive one-loop contributions to $\gtwo$, which arise from the 1-loop graphs in Figure~\ref{fig:diagrams} can become larger than the two-loop terms which are dominant in the first region. 

The third region represents the case where two-loop terms dominate, with a scalar mass in between the previous two cases. This is qualitatively somewhat similar to the first region with the roles of $A$ and $H$ reversed. However, this possibility requires a larger mass splitting than in the light $A$ scenario for two reasons.  One is that the new light scalar would need SM-like coupling to the top to generate a sufficient positive contribution, and thus the limits from $gg \to \phi \to \tau^+ \tau^-$ searches 
are stronger. The second is that the two-loop diagrams with a tau give a negative contribution when the scalar is lighter. This can be overcome by the top loops but a larger splitting is required.  With these constraints, and the upper limit on mass splitting imposed by bounds on $|Z_5|$, $A$ is required to fall in a $550-630$ GeV window with $m_H$ ranging from $30$ to $50$ GeV and $30 < |A_l| < 50$.

We note that of the flavor constraints listed in Eq.~(\ref{eq:flavor}), only  the first, $|A_l|/m_{H^+} < 0.4$ is relevant after applying our other requirements. As discussed in Section~\ref{sec:constraints}, this arises from universality tests in lepton decays and we replace this simple limit with the more complete calculation involving one-loop effects and measurements of several ratios, c.f. Eq.~(\ref{eq:hfag}). Qualitatively, one can see in Fig.~\ref{fig:zrange} that it is most important for light pseudo-scalars when $m_{H} \sim m_{H^+} \sim 200$ GeV, $m_A \gtrsim 100$ GeV and $|A_l| \gtrsim 60$. 

We compare $|A_l|$ and $|A_u|$ in the lower left panel in Fig.~\ref{fig:zrange}. For the light pseudo-scalars, with $m_A \sim 100$ GeV, $|A_l|$ and $|A_u|$ are anti-correlated. This is because, for fixed masses, high values of $|A_l|$ require a suppressed $|A_u|$ to avoid the upper $R_{B_s \to \mu^+ \mu^-}$ bound, while lower values of $|A_l|$ require $|A_u| \sim 1$ to give a sufficiently large contribution to $\Delta a_\mu$. As $m_A$ is lowered,  smaller values of $|A_u|$ are favored to avoid bounds from Higgs searches with the low mass accounting for the enhanced $\gtwo$ instead.
For sufficiently small $|A_u|$ however, tau loops dominate and constraints from direct production, $B_s \to \mu^+ \mu^-$ and $b \to s \gamma$ become unimportant. In the lower right panel of Fig.~\ref{fig:zrange} we plot $|A_u|$
against $|A_d|$. There is an upper bound which is driven by the constraint from $b \to s \gamma$. 
For very light scalars, couplings become constrained by searches for rare $\Upsilon$ decays to $\gamma \mu^+ \mu^-$ and $\gamma \tau^+ \tau^-$. Since light scalars will decay almost entirely to these leptons, this puts a strong bound on their coupling to $b$-quarks, requiring $|A_d| < 1$.
One can also see from these plots that the Type-X model limit, where $|A_d|=|A_u| = |A_l|^{-1}$, is only a small fraction of the viable A2HDM 
parameter space.

\subsection{$\gtwo$ in the CP-violating A2HDM}

New sources of CP violation, which are generically possible in the A2HDM, 
will give rise to new electric dipole moment (EDM) contributions for the electron and the neutron. Leading contributions to these 
dipoles are generated by Barr-Zee-type graphs which are directly analogous to those which are required to explain $\Delta a_\mu$. Unless we have a CP-preserving symmetry which ensures that the EDM contributions cancel, they will typically be too large to 
agree with the experimental limit. In the CP violating case with EDM constraints imposed, the largest $\Delta a_\mu$ 
generated in our general scan is an order of magnitude too small to account for the lower limit of the measured value \cite{cppaper}. 
Nonetheless, although the CP-conserving limit most naturally allows one to generate a large $\Delta a_\mu$ in the A2HDM, it is interesting to consider how far one can move away from this limit and still account for the experimental data.\footnote{See Ref.~\cite{cppaper} for details of the EDM calculation}

To investigate this we allow the parameters $\phi_u,~\phi_d,~\phi_l,~2\phi_5,~\phi_6$ and $\phi_7$ to deviate from a common phase $\phi_0$
\be
\phi_i = \phi_0 + \delta_i
\ee
where each $\delta_i$ is randomly distributed within a restricted range. For $|\delta_i| < 2\times 10^{-4}$ there is essentially no constraint from EDMs. The distribution of masses and parameter ranges is qualitatively the same as in the exact CP-conserving case. If we 
increase the allowed deviations to $|\delta_i| < 2\times 10^{-3}$, then the electron EDM becomes pertinent. Roughly two thirds of model points which would otherwise be allowed are excluded by the electron EDM \cite{cppaper}. At this level the neutron EDM predictions are still uniformly below the measured values. For $|\delta_i| < 2\times 10^{-2}$, approximately $98\%$ of the sample after other cuts is excluded by the electron EDM, and a few percent of the remaining points are now excluded by the neutron EDM.

Surviving points can still be found across the larger range of individual $\delta_i$, which suggests that sufficiently small EDMs can be accommodated by cancellations of several terms even for relatively large CP-violating phases. However, these canceling conditions become increasingly rarefied as the CP-violating phases grow. Without some principle which enforces such cancellations, we regard them as highly fine-tuned. Thus, while an explanation for the observed $\gtwo$ in the A2HDM does not strictly rule out significant new sources of CP-violation, they appear highly unnatural in this model. We will leave further discussions to a later publication \cite{cppaper}.


\begin{figure}
\begin{centering}
\includegraphics[width=8.1cm]{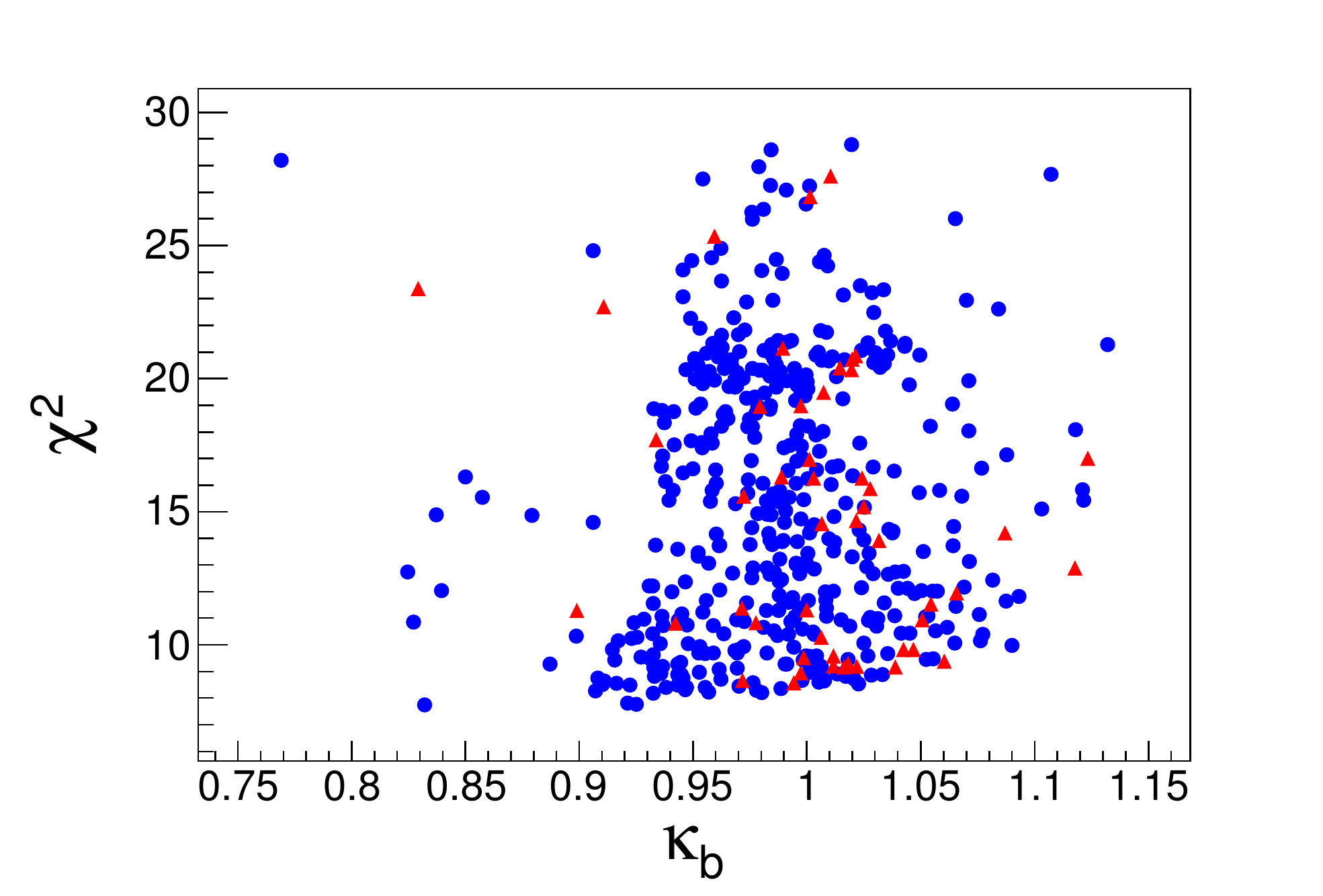}
\includegraphics[width=8.1cm]{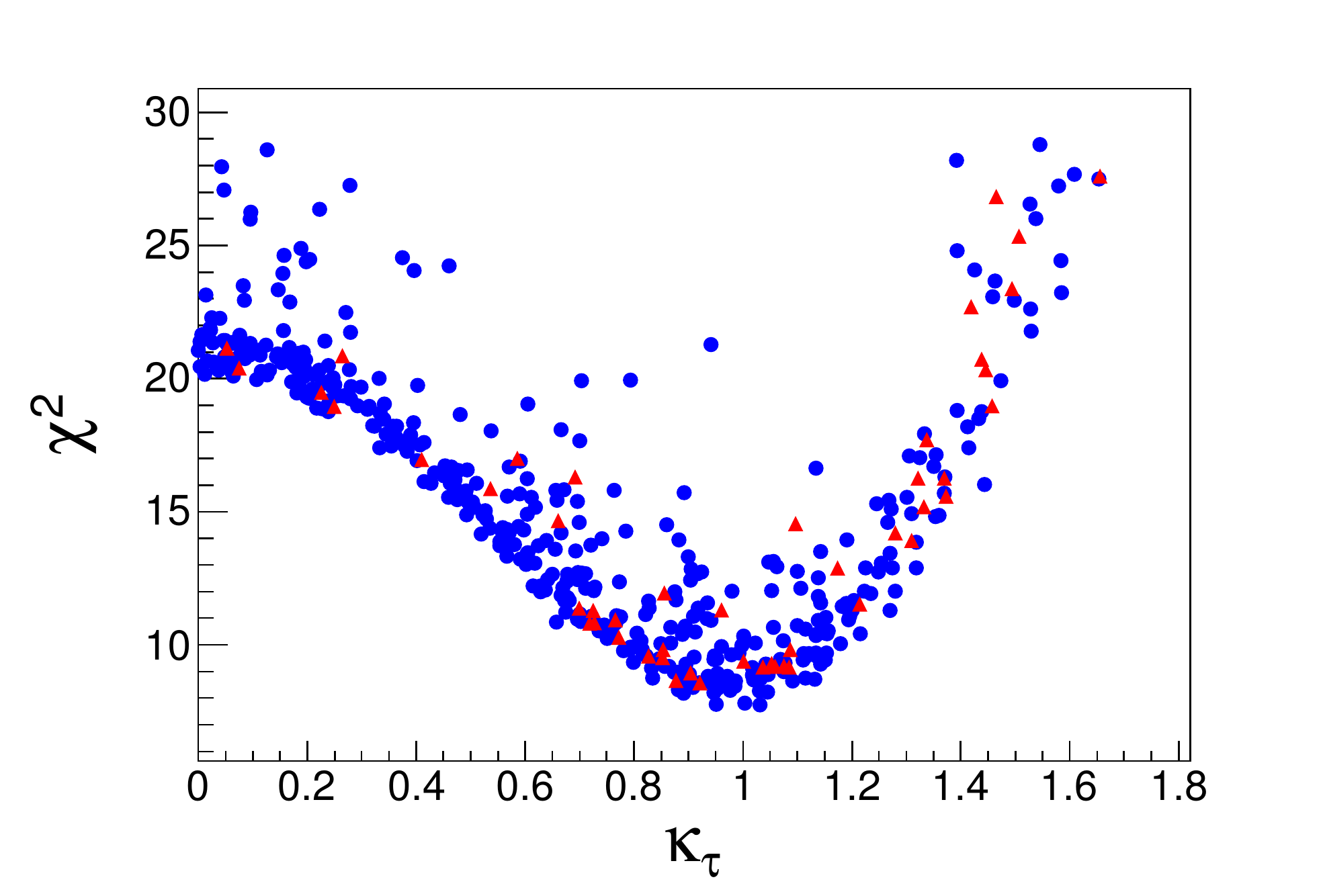}
\includegraphics[width=8.1cm]{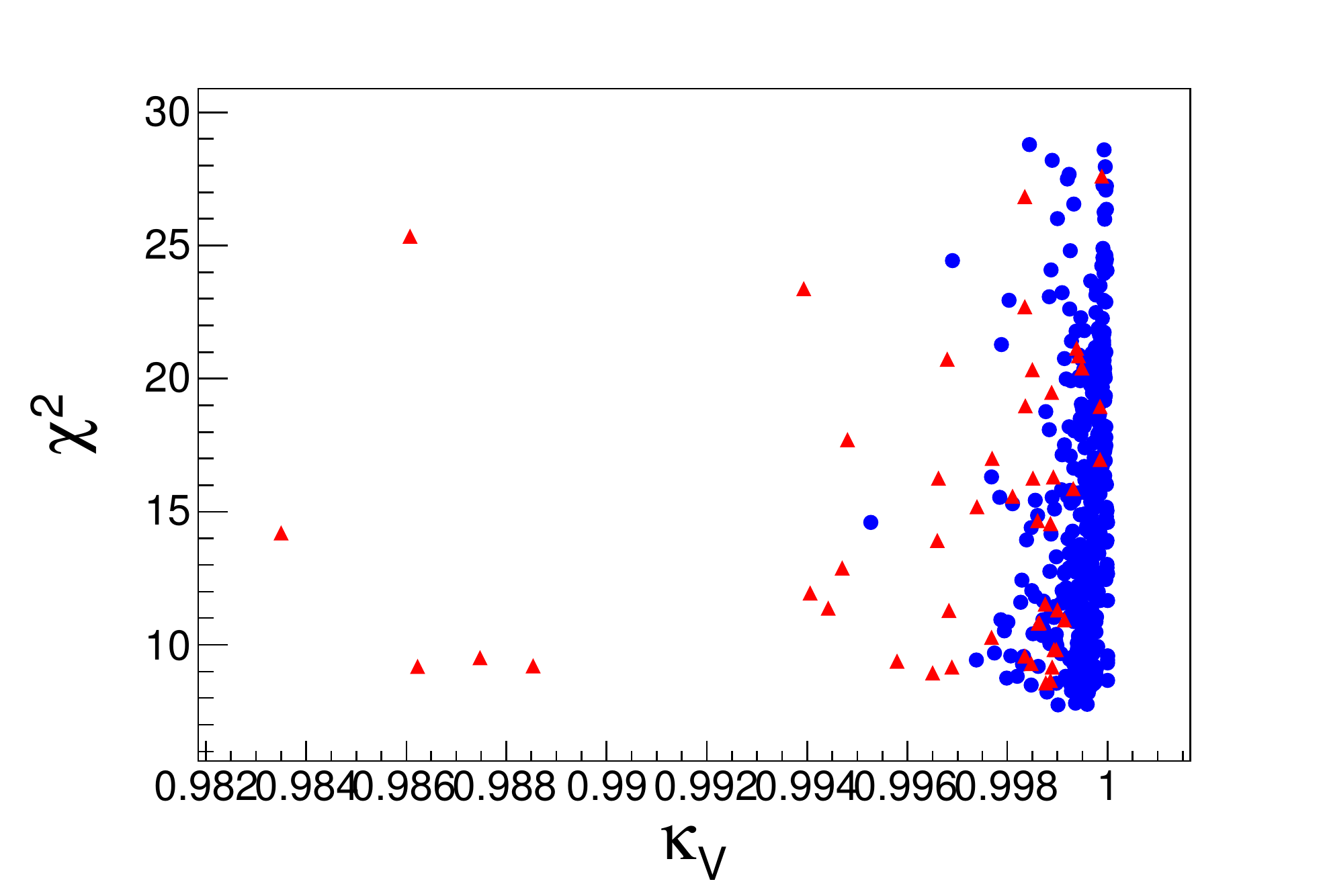}
\includegraphics[width=8.1cm]{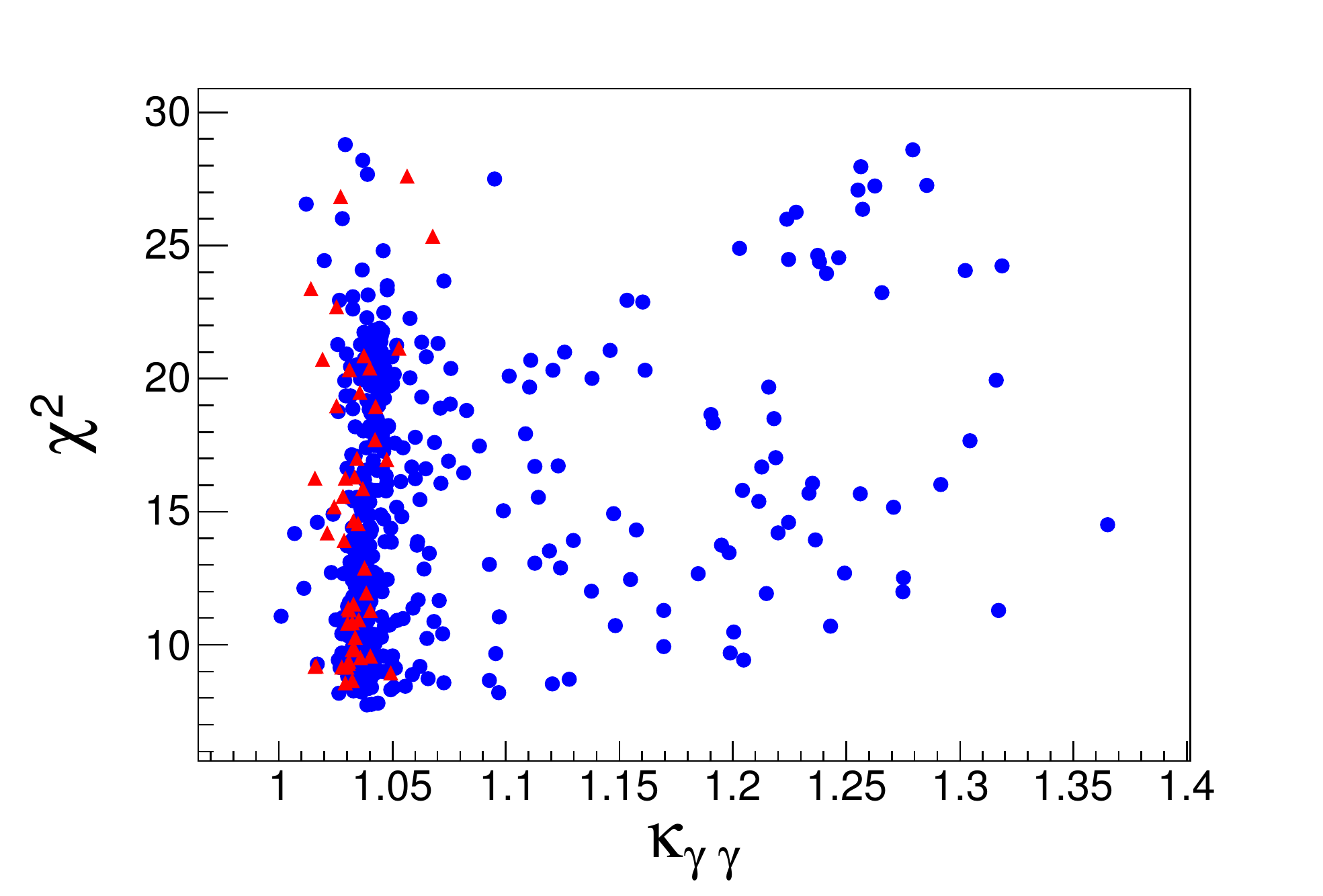}
\includegraphics[width=8.1cm]{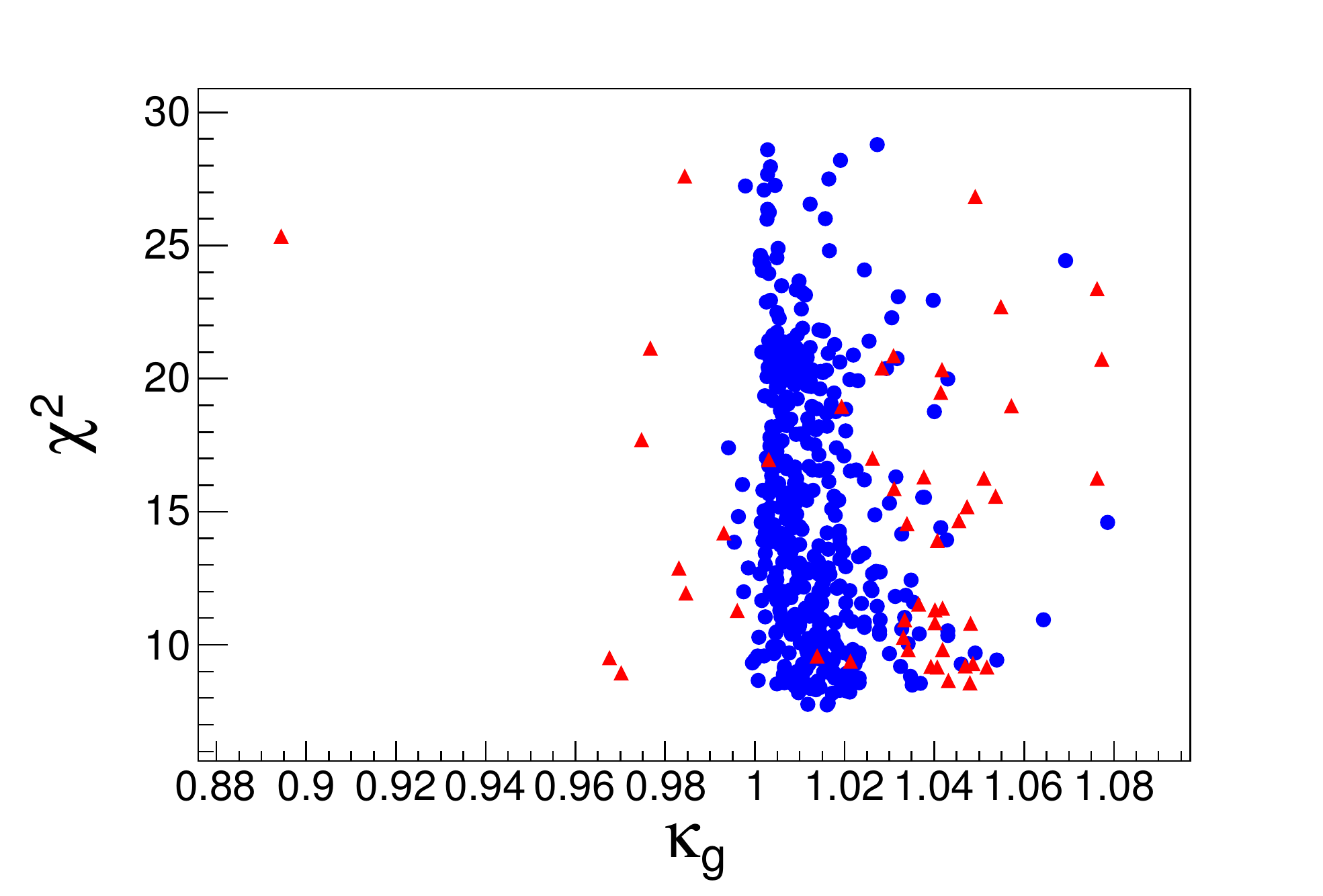}
\includegraphics[width=8.1cm]{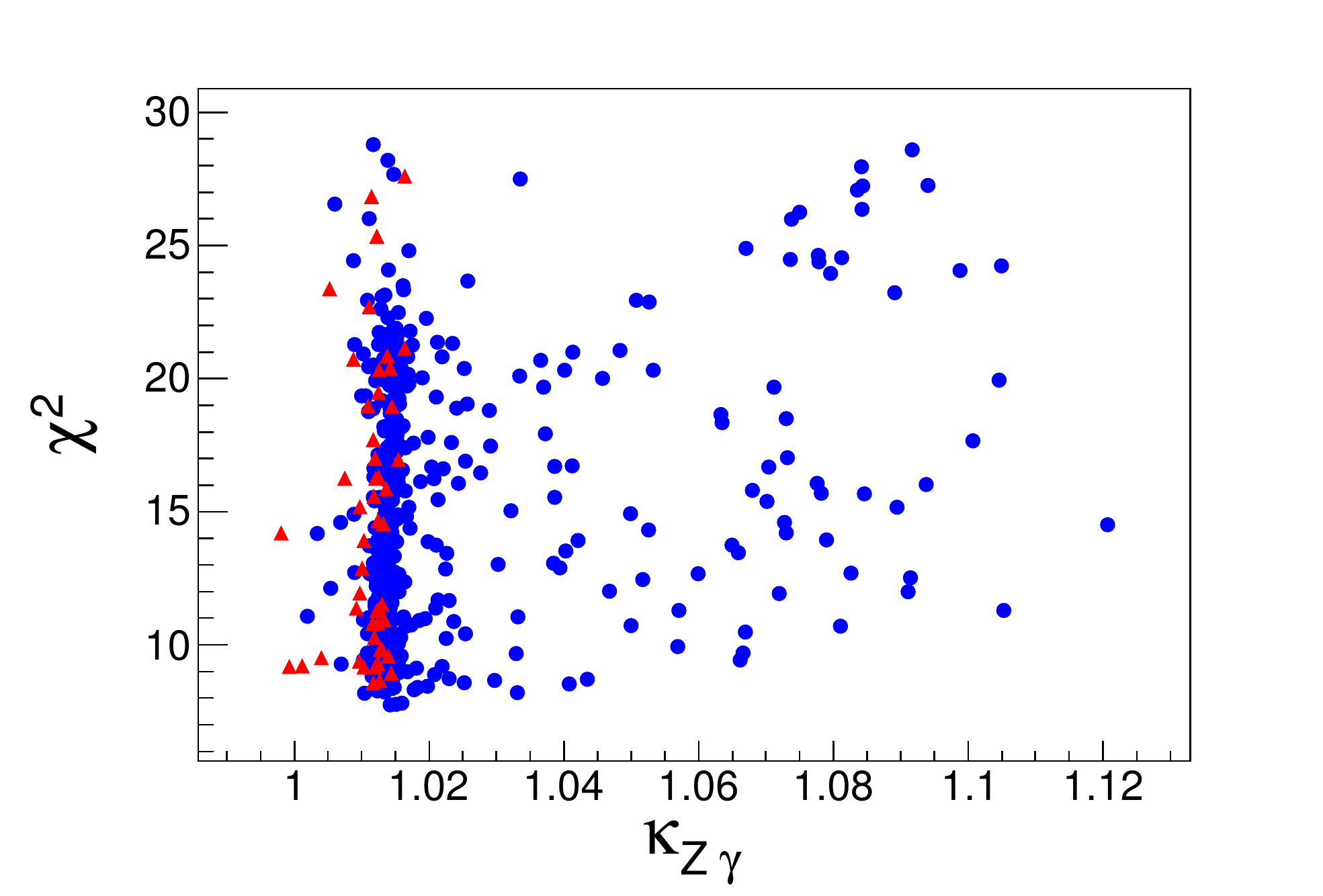}
\caption{Distribution of signal strengths versus $\chi^2$. From left to right at top: $\kappa_b$, $\kappa_\tau$; at middle:$\kappa_V$, $\kappa_{\gamma \gamma}$; at bottom: $\kappa_{g}$, $\kappa_{Z \gamma }$}
\label{fig:kappa_nocp}
\end{centering}
\end{figure}

\section{Modifications of SM Observables}
\label{sec:SM}

\subsection{Deviations of the SM-like Higgs couplings}
\label{sec:SMi}

The solutions to the $\gtwo$ excess in the A2HDM rely on the relatively light new neutral  states, which in turn modify certain SM observables. We first consider modifications to the SM Higgs properties at the observed $125$ GeV resonance.
The range of predicted coupling strengths $\kappa_i$, defined by $\kappa_i \equiv g_i/g_i^{SM}$ is shown in Fig.~\ref{fig:kappa_nocp} compared with the $\chi^2$ fit. As expected, deviations are driven by the large values of $|A_l|$, which are apparent in the potentially large modifications of $\kappa_\tau$. One can see that $\kappa_\tau$ may even vanish without violating our bound on the combined fit to data, although this is disfavored by the $\tau\tau$ channel itself. The fit to LHC data 
directly limits these excursions, which in turn requires that the mixing angle $\theta_{12}$ is small. This mixing determines $\kappa_V$, which is thus constrained to be very close to $1$.  We find $\kappa_V > 0.998$, except for the case of very light scalars, which may accomodate $\kappa_V > 0.982$. In either case, this is much more restrictive than the current or even expected limits from direct fits to $\kappa_V$. This is equivalent to a mixing angle between the two scalar states of $\sin \theta_{12} <0.06$ ($0.19$ for very light scalars.)  Since mixing is constrained to be so small and since the enhancement of other Yukawa couplings cannot be as large as $|A_l|$, we don't see large deviations in the other tree level couplings. In general $\kappa_b$ can vary by about $10\%$ relative to the Standard Model. The effective coupling to $\gamma \gamma$ and to $\gamma Z$ is typically enhanced by a few percent, although the former can be as large as $40\%$ high and the latter as large as $10\%$ high. In the case of light scalars only $1-5\%$ deviations are expected for the $\gamma \gamma$ and $\gamma Z$ couplings. The coupling to gluons is predicted to be within a few percent of the Standard Model value with $\sim 10 \%$ deviations possible with light scalars. 

\begin{figure}
\begin{centering}
\includegraphics[width=8cm]{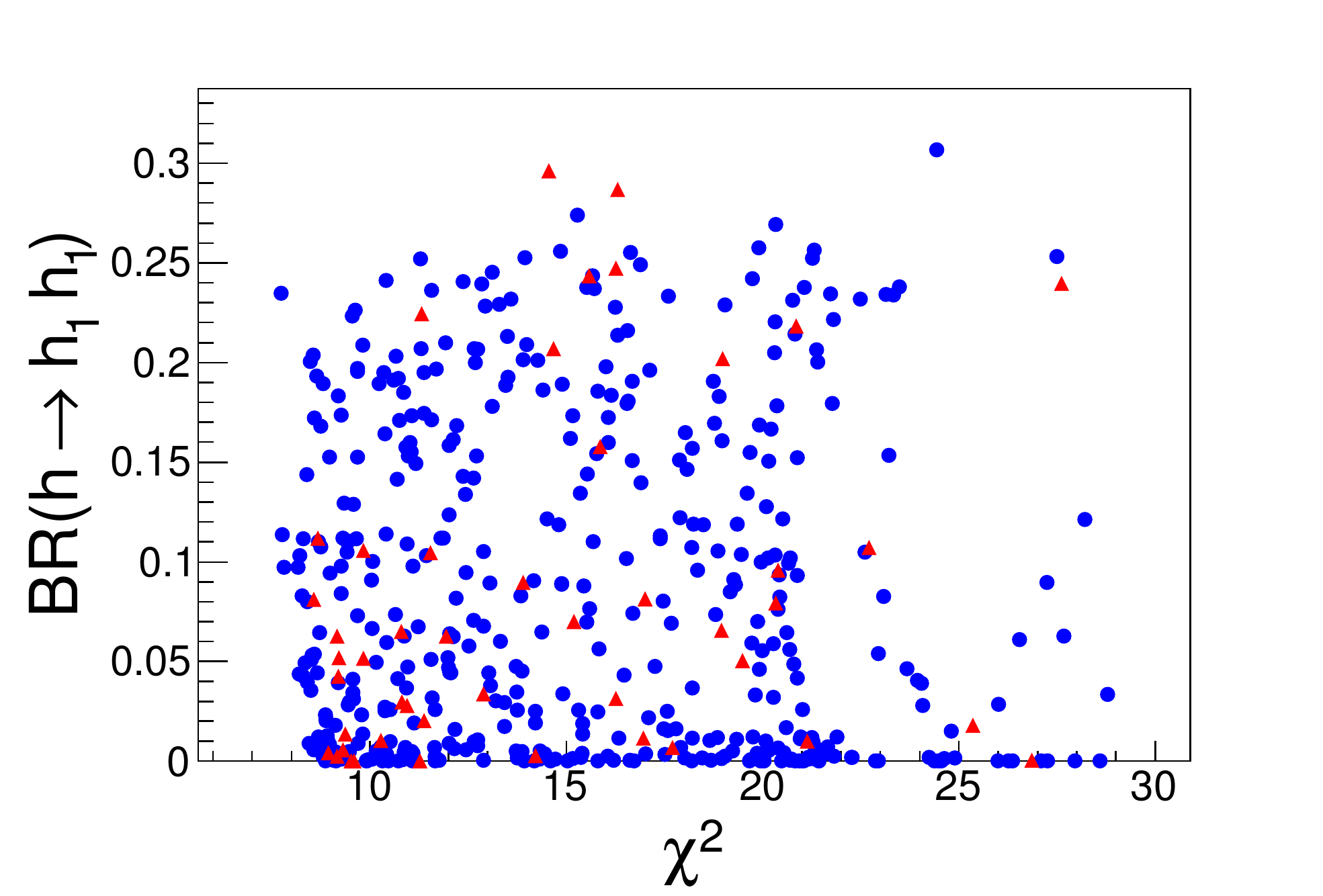}
\includegraphics[width=8cm]{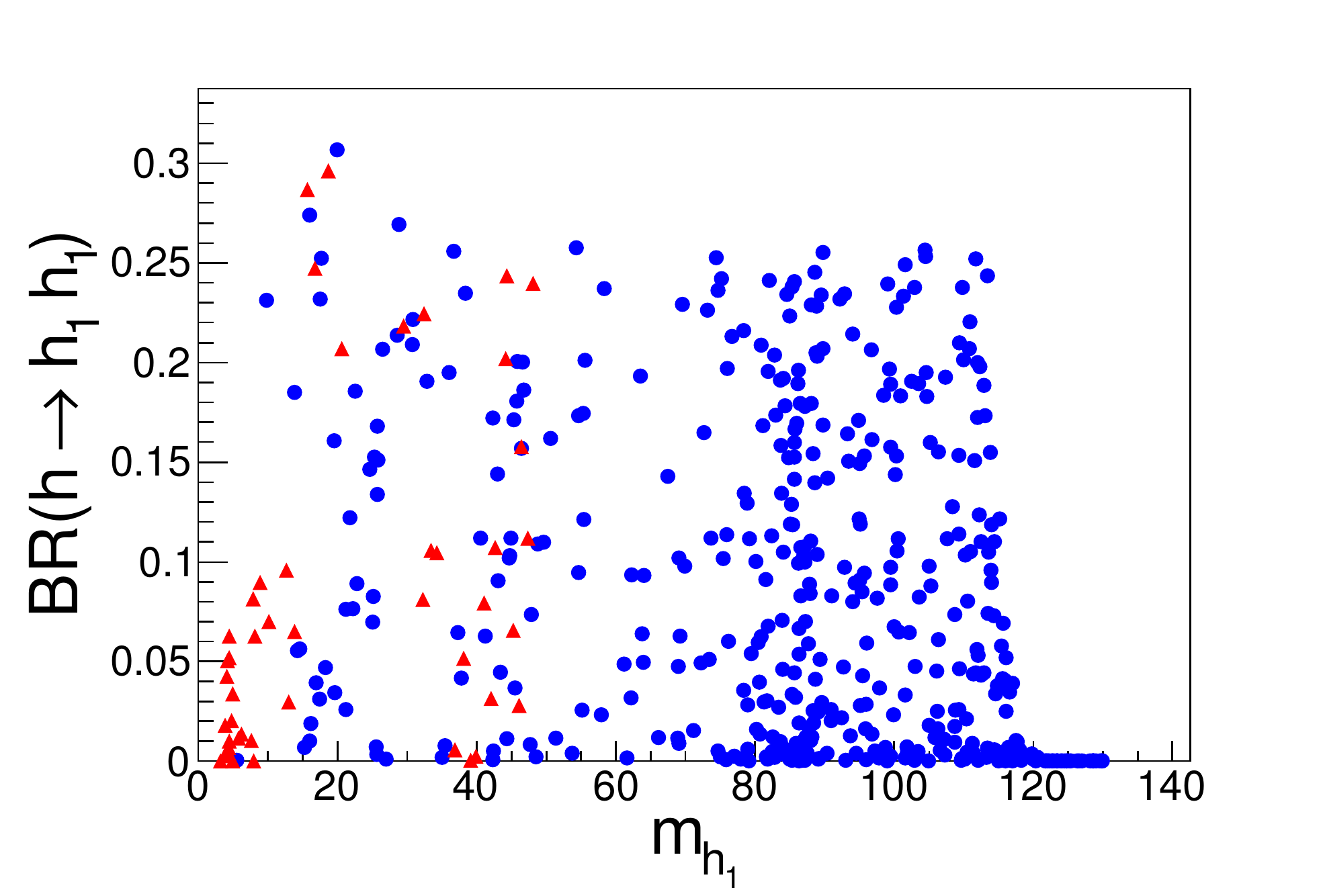}
\caption{
Branching fraction for the SM-like Higgs boson to two light Higgs states versus the fitted $\chi^2$ (left panel) and $m_{h_1}$ (right panel). 
}
\label{fig:BRH}
\end{centering}
\end{figure}

With these relatively mild deviations, the variance in $\chi^2$ values strongly  tracks the fit to $\tau \overline{\tau}$ data.
The best fit value is $\chi^2 \simeq 8$, well within the $1\sigma$ expectation for the number of degrees of freedom (18). For comparison the Standard Model value in our fit is $\chi^2_{SM} \simeq 13$. 
The total width of the SM-like Higgs boson can differ in our results in the range $3.5 - 7$ MeV. The upper range comes from decays to pseudo-scalar pairs. As discussed above, we limit these to a $25\%$ branching fraction based on three-lepton searches for $h_1 > 30$ GeV and lighter masses are constrained by published results. The $\chi^2$ fit by itself would rule out values higher than $\sim 50\%$.

Figure \ref{fig:BRH} shows the achievable branching fraction for the SM-like Higgs boson decay to two light Higgs states versus the fitted $\chi^2$ (left panel), and $m_{h_1}$ (right panel). 
We note that, for $m_{h_1} < 115$ GeV, a significant branching fraction is possible for the SM-like Higgs boson $h$ to decay via two $A$'s or two $H$'s into four taus. This can be as high as the imposed limit of $25\%$ for $m_A \sim 110$ GeV, even though the off-shell $A$ suppression is strong, due to the possibly large values of the $hAA$ coupling and $|A_l|$. It is thus strongly motivated to search for the exotic decay of the SM-like Higgs boson to $4\tau$'s at the LHC experiments. For low masses of $h_1$ the branching ratio is restricted by experimental searches, particularly by 
$h \to AA \to  2\tau 2\mu$. Below the $2\tau$ threshold muon decays tend to dominate and only a very small branching fractions of $h \to AA$ is allowed, so that $m_{h_1} < 3$ GeV does not appear in our scan. However, it is typically possible to arrange a partial cancellation of the terms which contribute to the $hAA$ coupling, leading to arbitrarily small branching fractions. Thus it is possible to allow for  lighter $h_1$ masses down to $\sim 1$ GeV if this coupling is set sufficiently small.

\subsection{Precision Observables in the SM}
\label{sec:SMii}

\begin{figure}
\begin{centering}
\includegraphics[width=8cm]{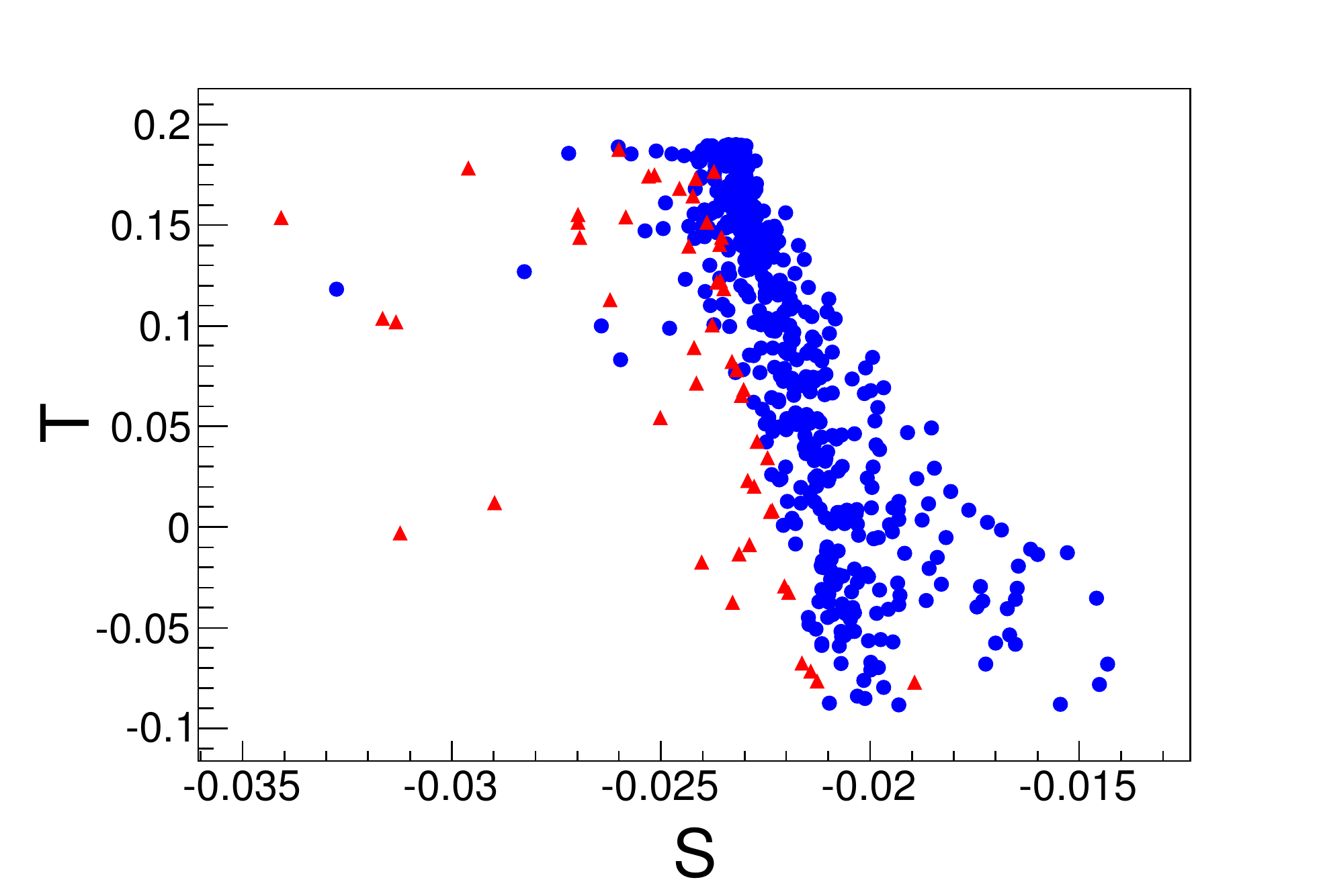}
\includegraphics[width=8cm]{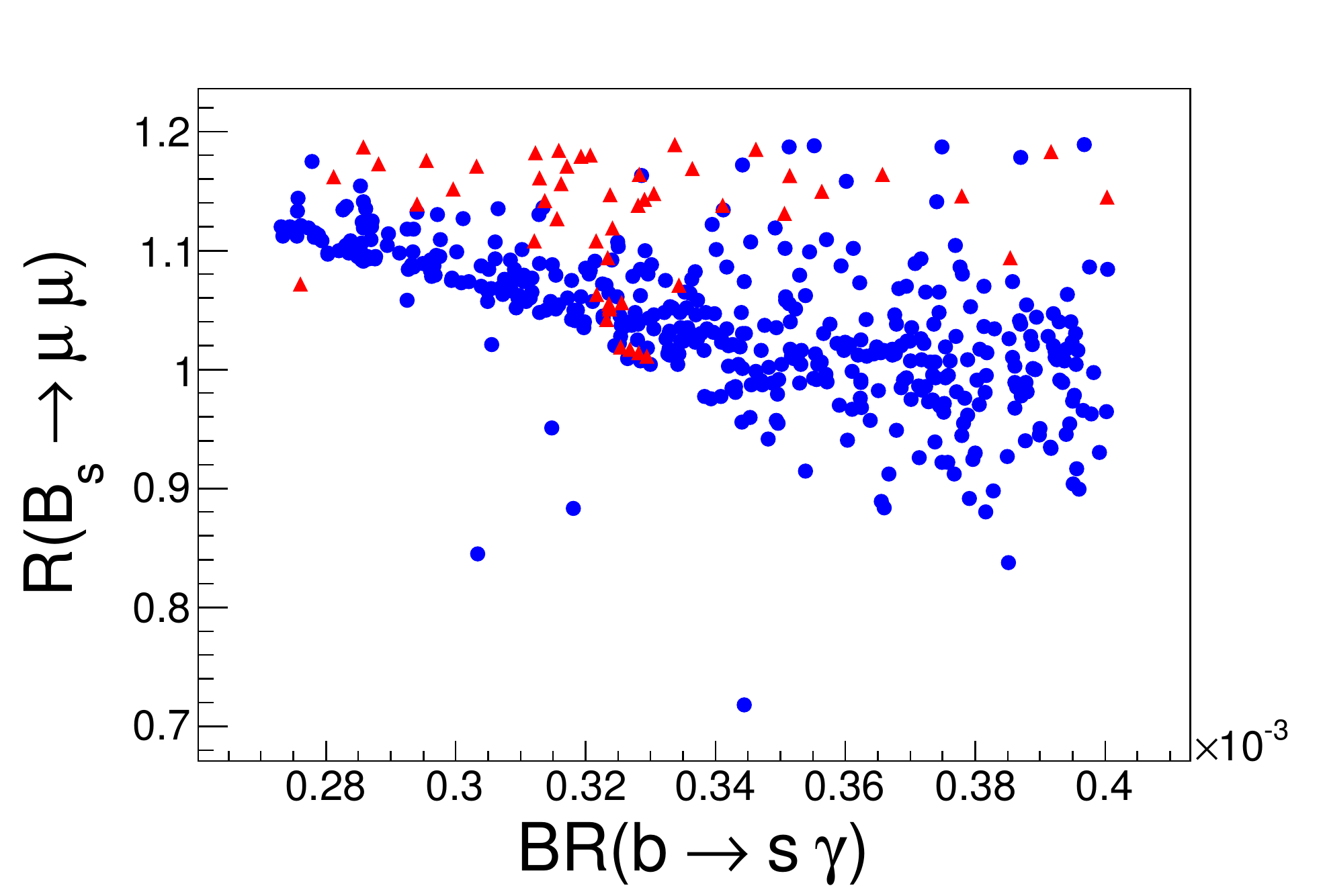}
\caption{Left: Range of the S and T parameters predicted in the CP-conserving A2HDM. Right:  Comparison of predicted values for $BR(b \to s \gamma)$ and $R_{B_s \to \mu^+ \mu^-}$.}
\label{fig:stnocp}
\end{centering}
\end{figure}

In Fig.~\ref{fig:stnocp} we show predictions for the EW precision parameters $S$ and $T$ in the left panel, and the ratio $R_{B_s \to \mu^+ \mu^-}$ versus the branching fraction $BR(b \to s \gamma)$ on the right.  As in the CP-violating case the experimental value for $S$ is not a significant constraint, although it is predicted to have a negative shift relative to the SM value $S=0$. Predicted values of $U$ are 
 similarly within the experimental error, however, $T$ is strongly constraining from either direction. Turning to rare decays, $b \to s \gamma$ is a relevant constraint in general, although light scalars seem to prefer a central value.  $R_{B_s \to \mu^+ \mu^-}$  is also an important bound on the allowed space, with an interesting distribution. The central experimental value is currently about one standard deviation below the SM value, whereas most of the light pseudo-scalar and all of the light scalar points are above it.  If it were truly as low as $R< 0.85$ then only a few points with $m_H^0 \simeq m_H^+ \simeq 200$ GeV would survive. We also note that such points are relatively even  rarer in the CP-violating case without $\gtwo$ \cite{cppaper}.

\begin{figure}
\begin{centering}
\includegraphics[width=8cm]{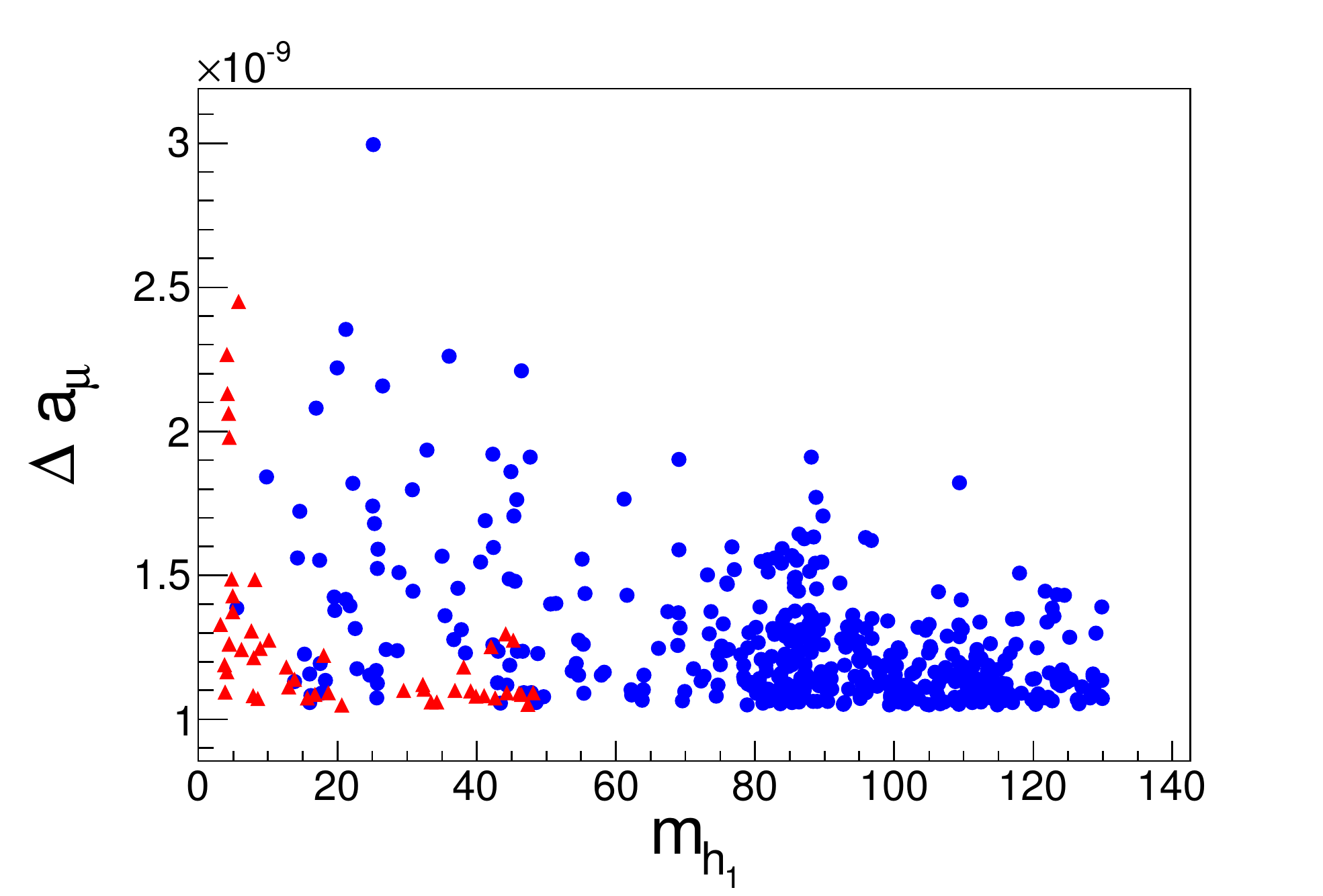}
\includegraphics[width=8cm]{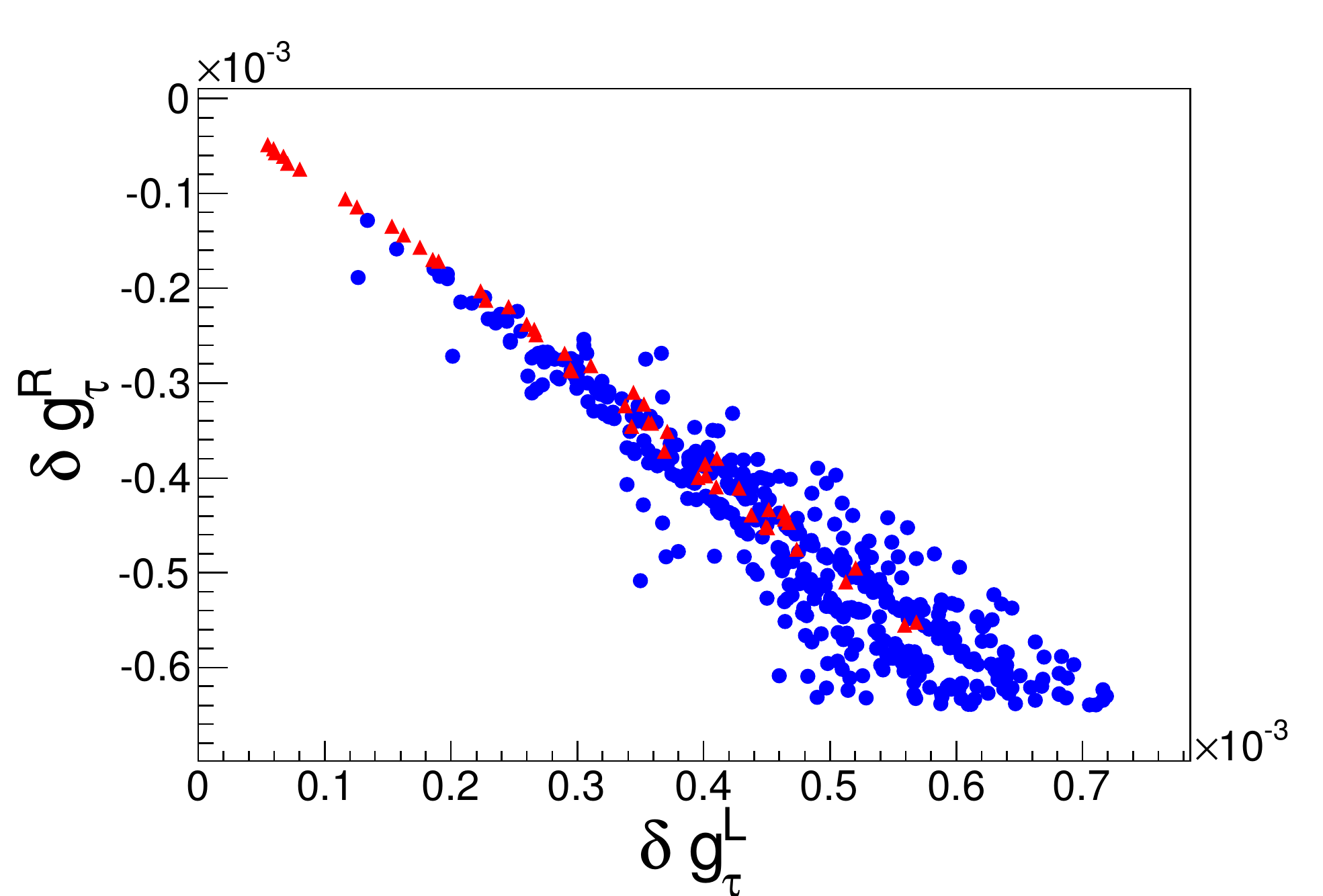}
\includegraphics[width=8cm]{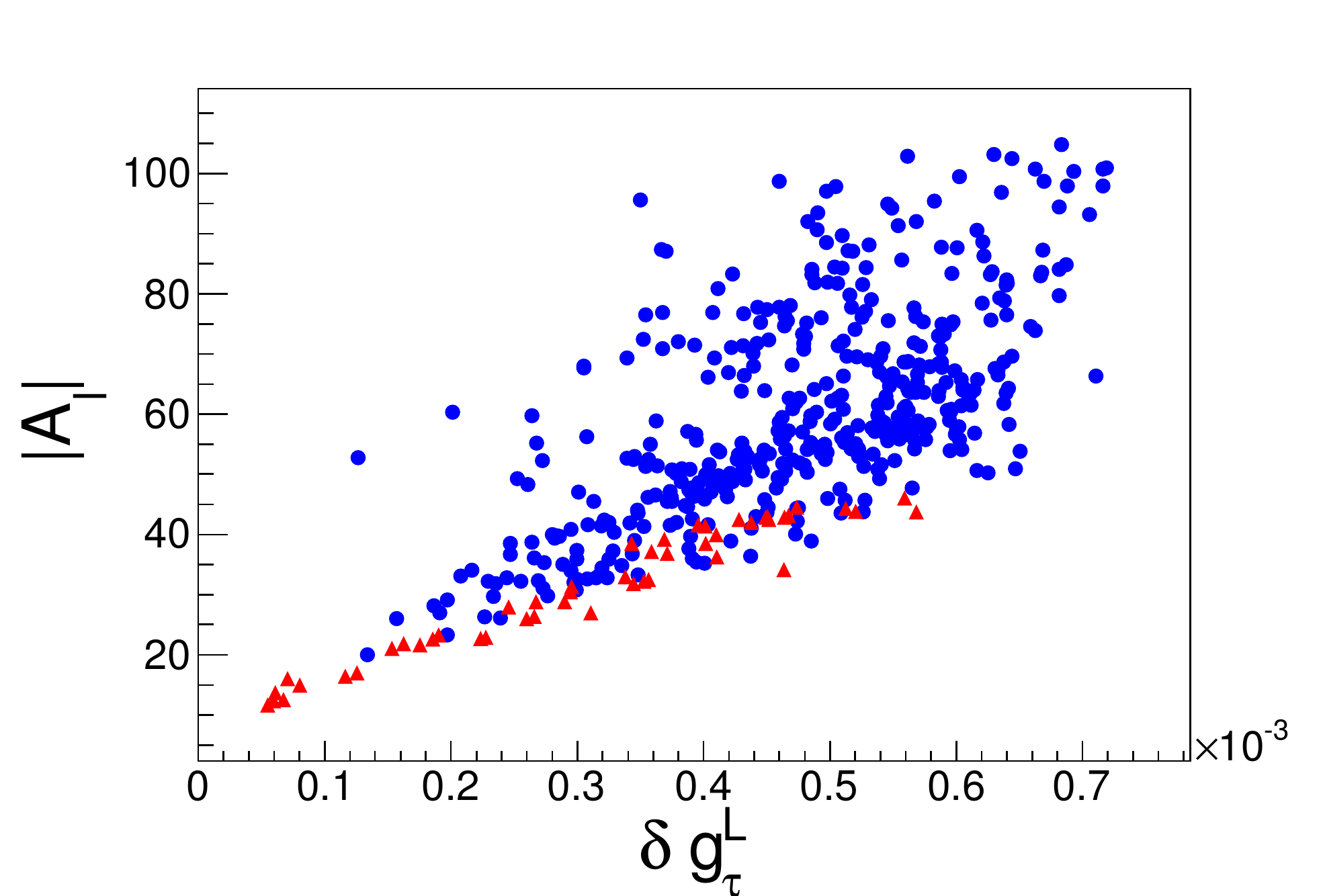}
\includegraphics[width=8cm]{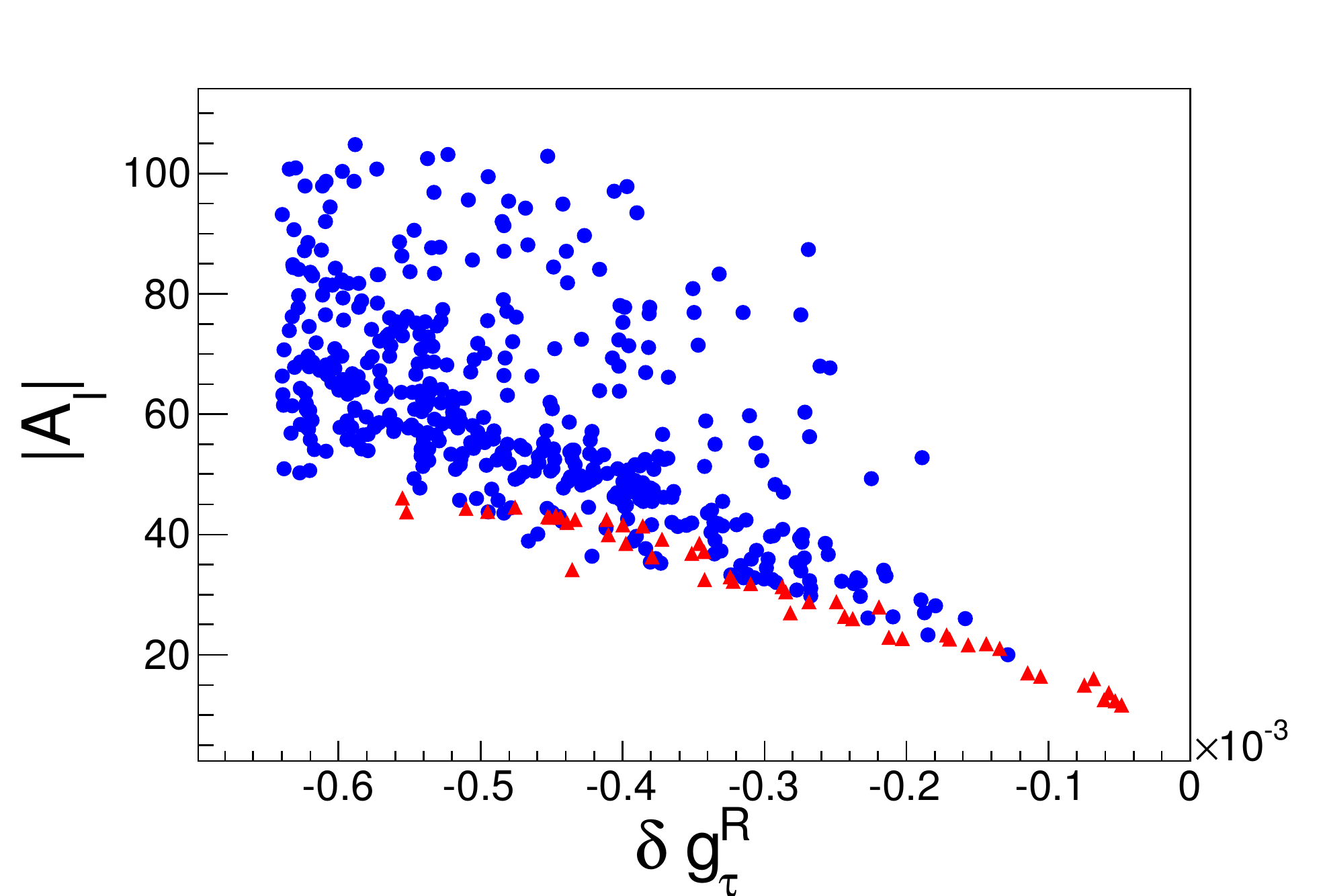}
\caption{Upper left: The predicted value of $\Delta a_{\mu}$ versus $m_{h_1}$; Upper right: Magnitude of the deviations from SM couplings of $Z\tau\tau$ of $\delta g_\tau^L$ versus $\delta g_\tau^R$;
Lower left: $\delta g_\tau^L$ versus $|A_l|$. Lower right:  $\delta g_\tau^R$ versus $|A_l|$.}
\label{fig:gmu}
\end{centering}
\end{figure}

As for $\Delta a_{\mu}$ itself, values at the lower bound of the experimental range are most common in our results.   Nonetheless, much of the range of experimentally consistent results can be explained by the model for the regions with  very light scalars or with light pseudo-scalars, although points above the central value are quite rare. We show the predicted value of $\Delta a_{\mu}$ versus the mass parameter $m_{h_1}$ after all constraints have been implemented in the upper-left panel of Fig.~\ref{fig:gmu}. The E989 experiment at Fermilab is expected to improve experimental errors on the $\gtwo$ measurement by a factor of four \cite{Grange:2015fou}.

In the upper-right panel of Fig.~\ref{fig:gmu},
we display the predicted values for $\delta g_\tau^{L,R}$. The magnitudes of the left and right-handed couplings are strongly correlated with $\delta g_\tau^L > 0$ and $\delta g_\tau^R < 0$. This is opposite to the small shifts from Standard Model predictions found experimentally. In particular, since $g_\tau^R$ is high by roughly one standard deviation, it provides an important constraint on our model. In the lower panels of Fig.~\ref{fig:gmu}, we display the predicted coupling shifts as a function of $|A_l|$. One can see that the measured value of $g_\tau^R$ becomes a significant bound for $|A_l| > 50$ and excludes a significant fraction of otherwise viable points. Since $\delta g_\tau^L$ is highly correlated with $g_\tau^R$ and 
the experimental fit for the latter is closer to the SM, $\delta g_\tau^L$ does not impose any relevant additional constraint.
%

\section{Searches for new Higgs states}
\label{sec:search}

Going beyond the direct search for the exotic decay of the SM-like Higgs boson $h\to h_1 h_1 \to 4\tau$ at the LHC experiments, and the indirect tests by measuring the Higgs couplings, precision EW parameters, as well as the $B$-meson rare decays, 
an important prospect is the potential discovery of new states at the LHC.
The lighter new neutral state will decay almost entirely to taus, except for the very light scalars in the second region, 
which may decay to taus, or to muons if they are below the tau-pair threshold. The heavier new state must typically be massive enough to decay to the lighter Higgs state or to tops in order to limit the decay to taus and to suppress negative contributions to $\gtwo$, which is strongly constrained by LHC searches.

\begin{figure}[tb!]
\begin{centering}
\includegraphics[width=8cm]{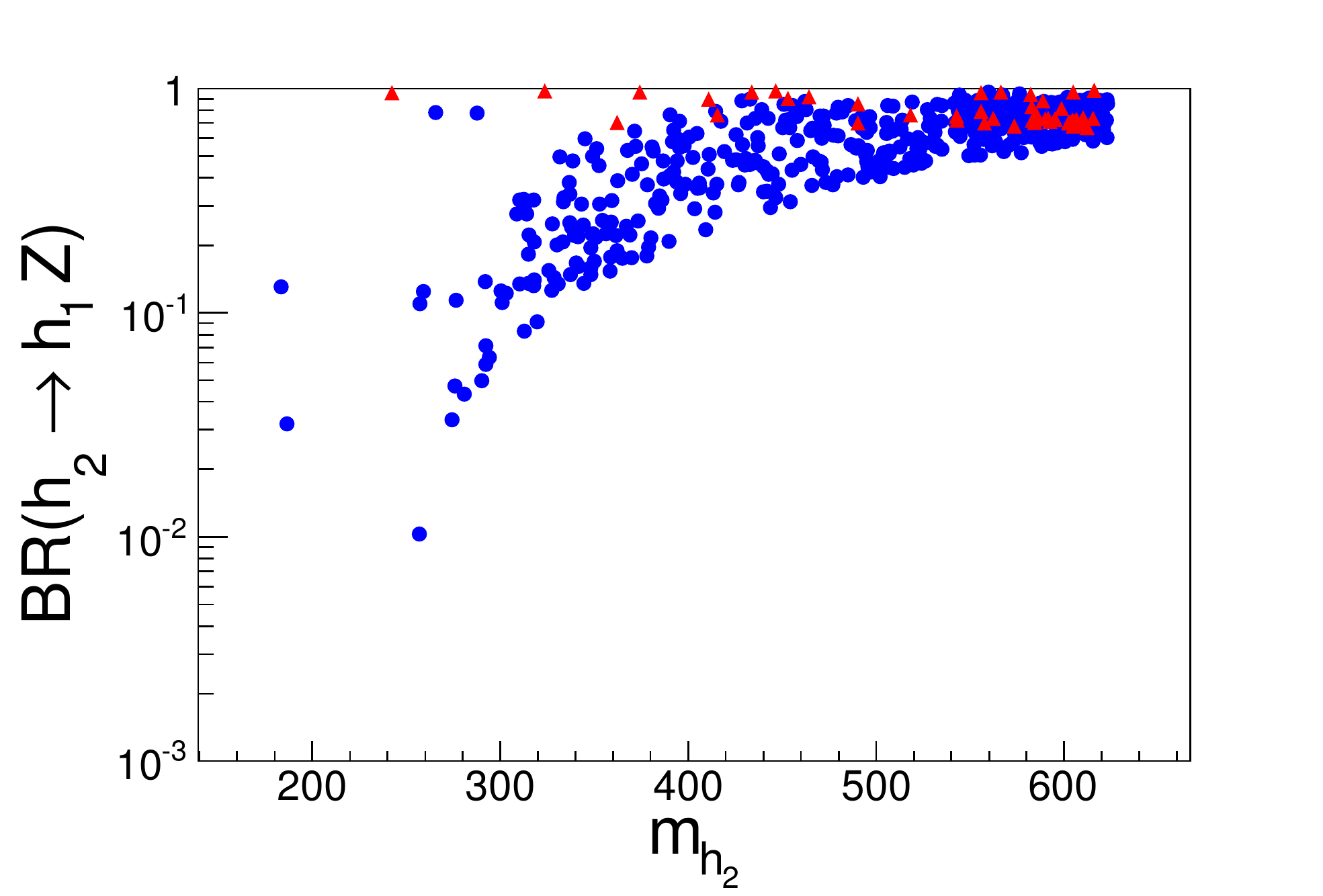}
\includegraphics[width=8cm]{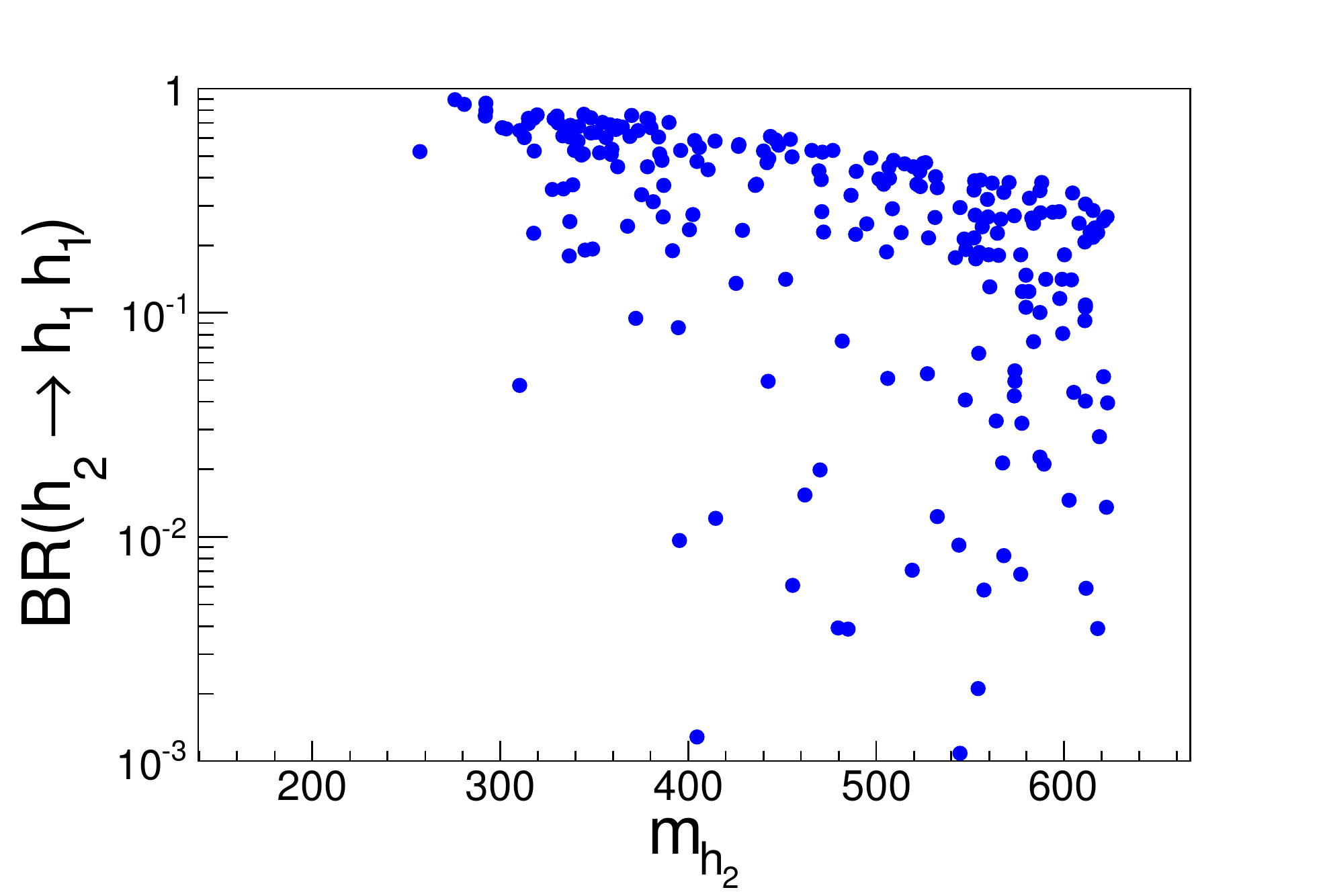}
\includegraphics[width=8cm]{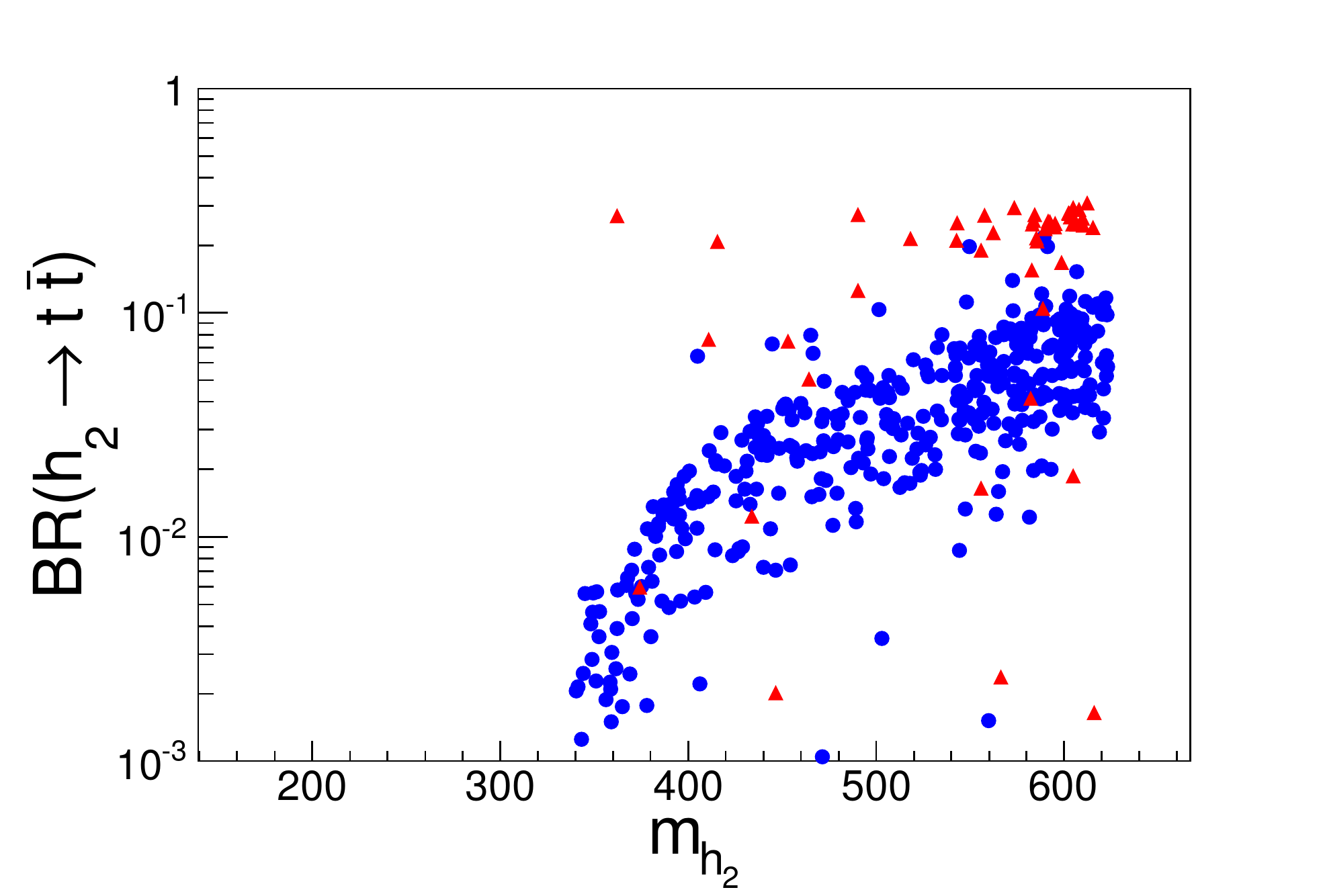}
\includegraphics[width=8cm]{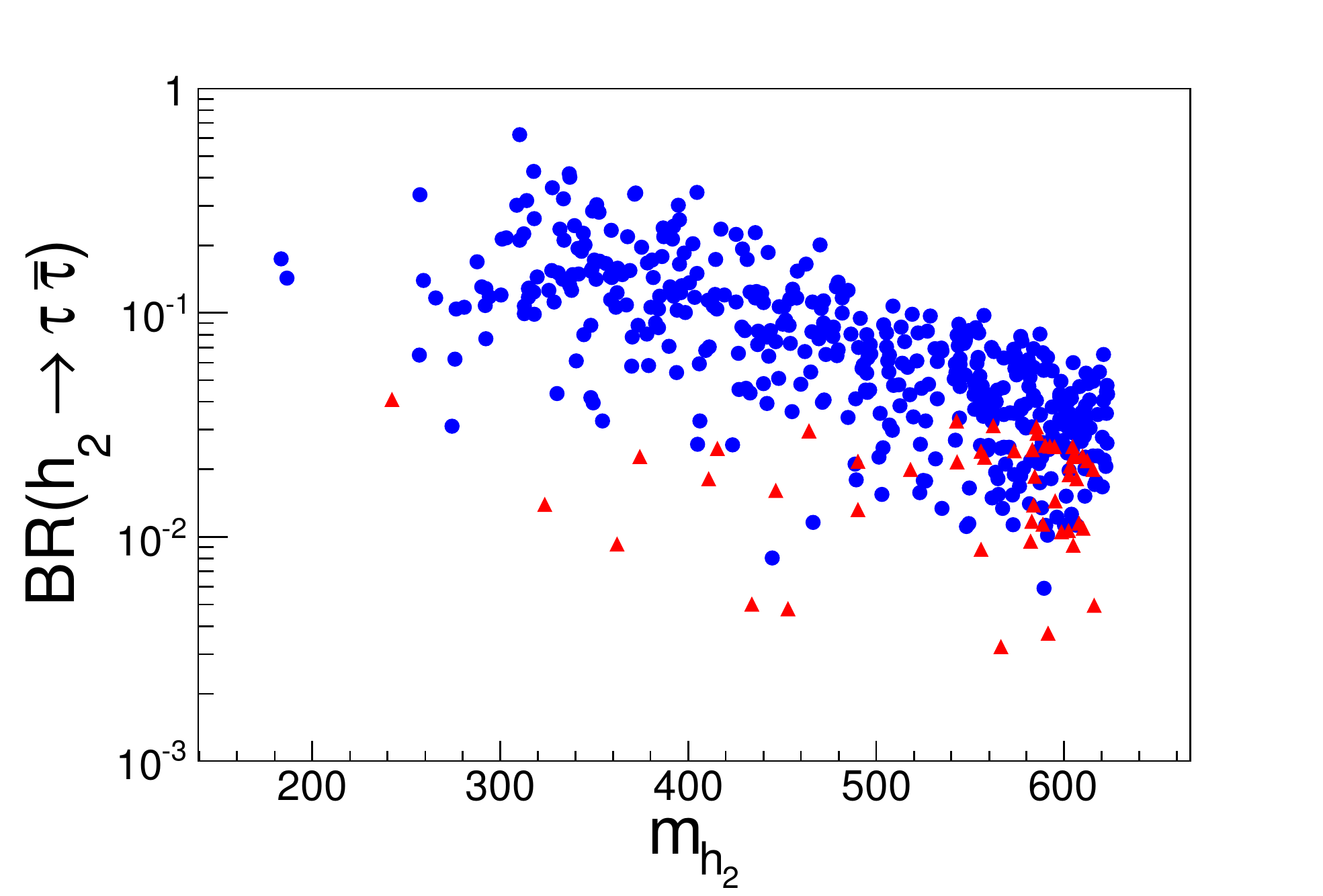}
\includegraphics[width=8cm]{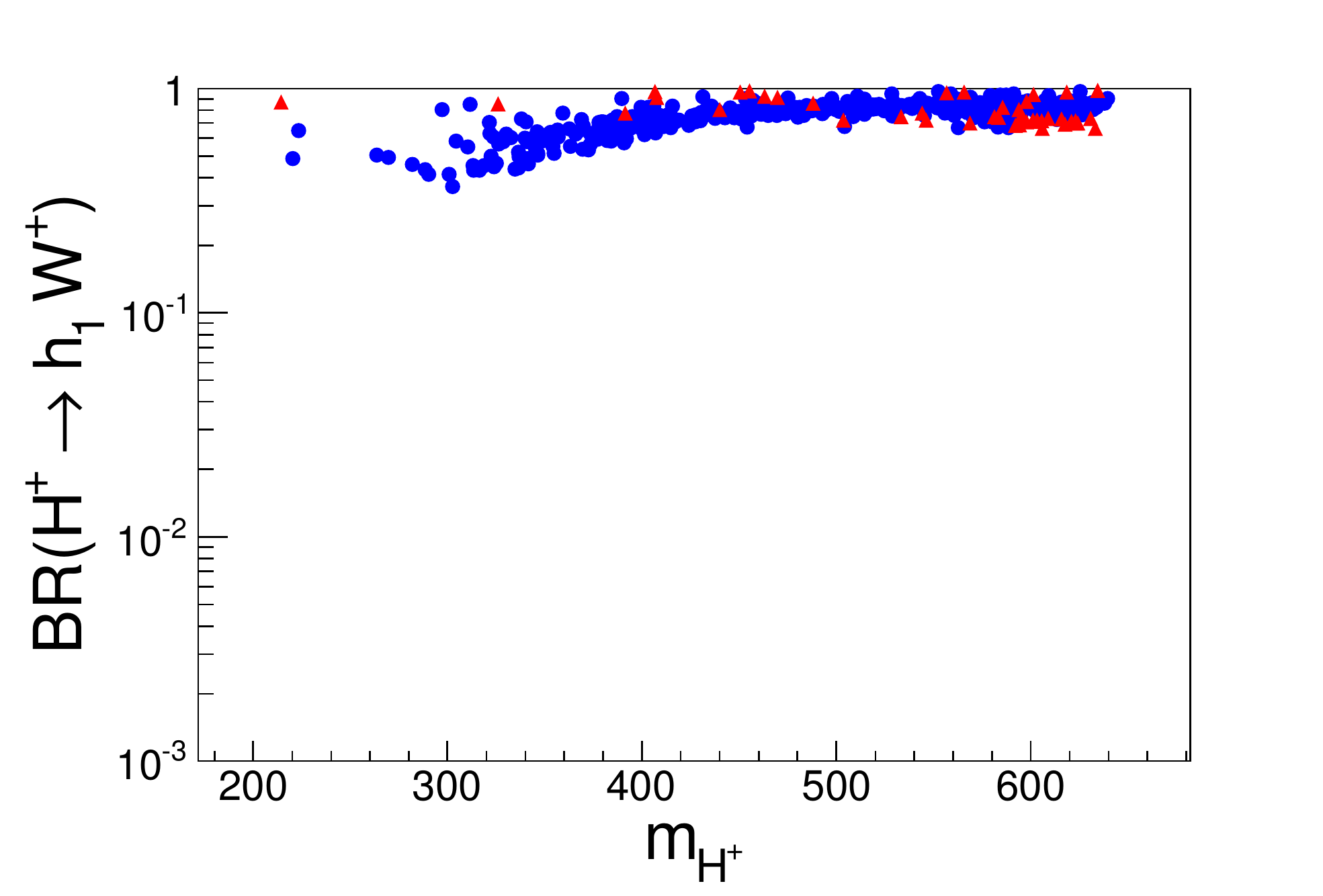}
\includegraphics[width=8cm]{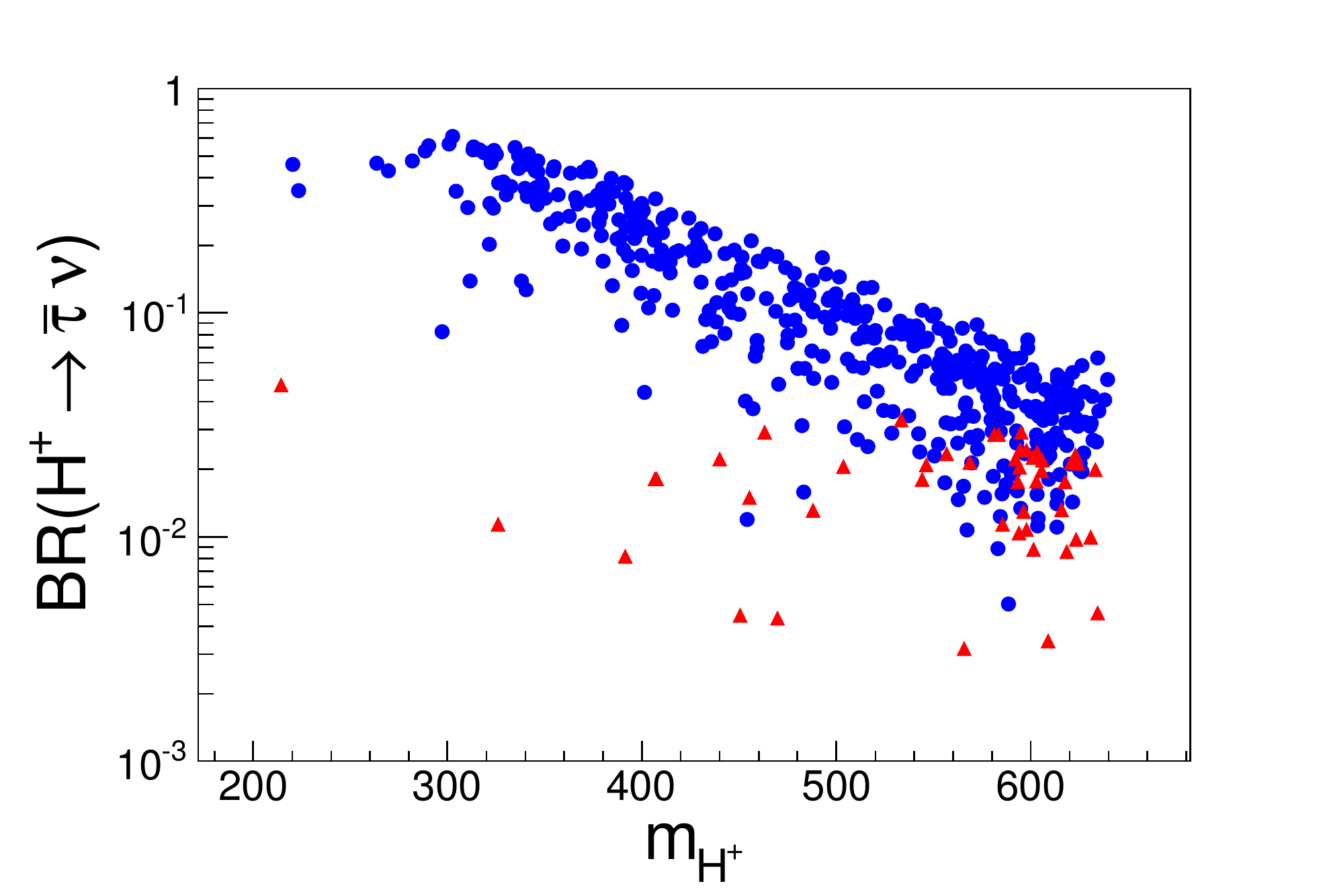}
\caption{Predicted branching fractions of heavy new states. 
At top :  decays of the heavier neutral Higgs $h_2\to h_1 Z$ on the left and to $h_1 h_1$ on the right, where $h_1$ is the lighter new neutral state.
At middle: decays of $h_2$ to tops on the left and to taus on the right. 
At bottom: decays of the charged Higgs $H^\pm\to h_1 W$ on the left and to $\tau \nu$ on the right.}
\label{fig:bratios}
\end{centering}
\end{figure}

In Fig.~\ref{fig:bratios} we present the leading branching fractions for the decay of the heavier neutral state ($h_2=H$ or $A$) and the charged Higgs ($H^\pm$). The decays of the heavier neutral state are shown in the top four panels in Fig.~\ref{fig:bratios}. 
The two leading, but complementary, channels are $h_2 \to h_1Z,\ h_1 h_1$,  
as shown in the two top panels of the figure.
For all cases we find that the heavier new neutral state may decay to tops with at most $30 \%$ branching fraction and to taus with at most $50 \%$, as seen in the two middle panels. 
In general the maximum branching fraction to taus or to $h_1 h_1$ declines at higher masses, while the fraction for decay to $Zh_1\ (H\ {\rm or}\ A)$ increases. 
Decays to $b\overline{b}$ are relatively rare since only a mild enhancement over SM-like couplings is allowed.
As seen in the two lower panels in Fig.~\ref{fig:bratios},
the charged Higgs can have a large branching fraction to $\tau \nu$ if $|A_l|$ is large which tends to dominate at lower $m_{H^+}$ with 
light pseudo-scalars. At higher masses, and generally in the case of light scalars, the charged Higgs will decay primarily to $W h_1$. After accounting for these two decays any remaining fraction will be largely $H^+ \to t \bar{b}$, which may reach about $20 \%$.

\begin{figure}[tb!]
\begin{centering}
\includegraphics[width=8cm]{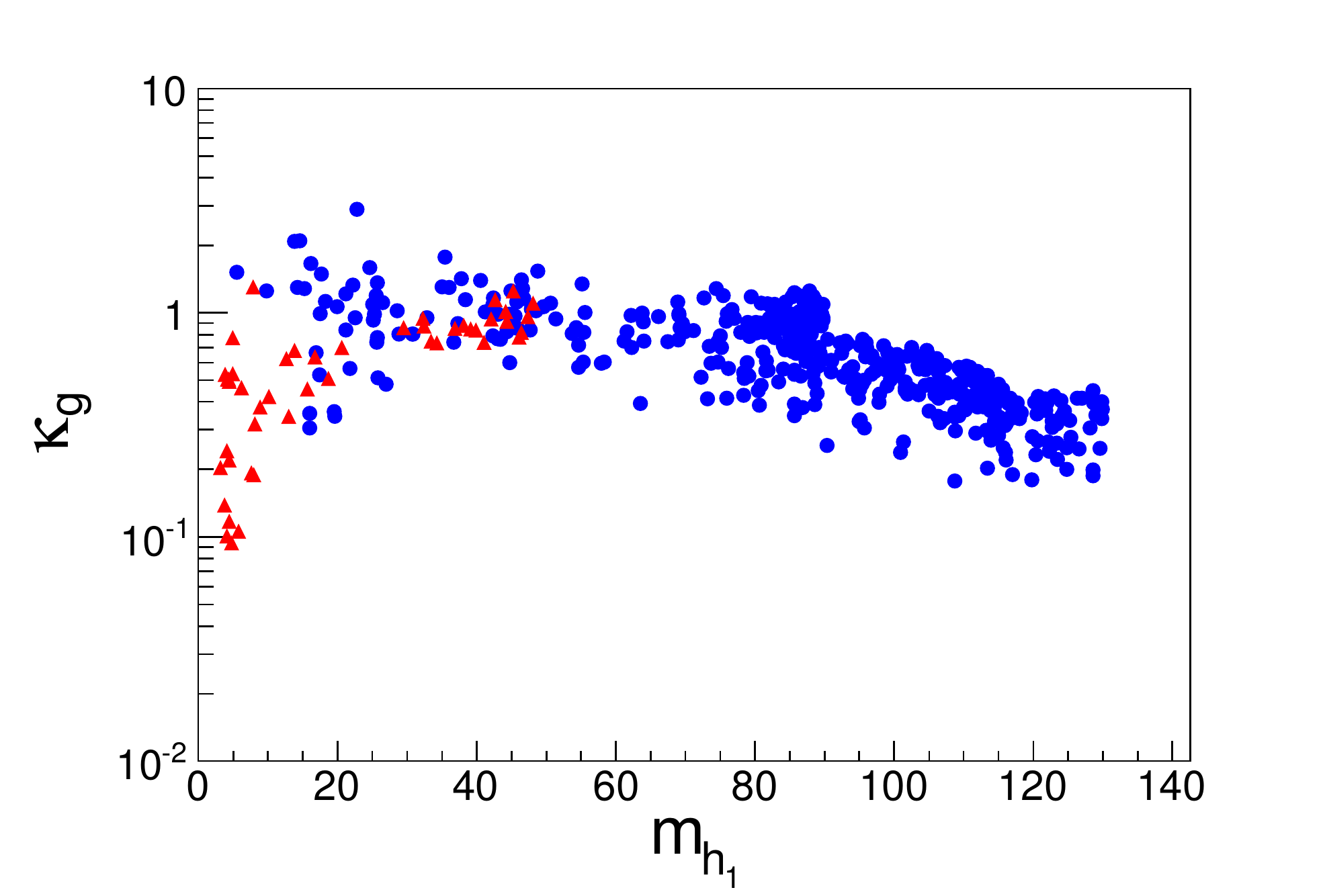}
\includegraphics[width=8cm]{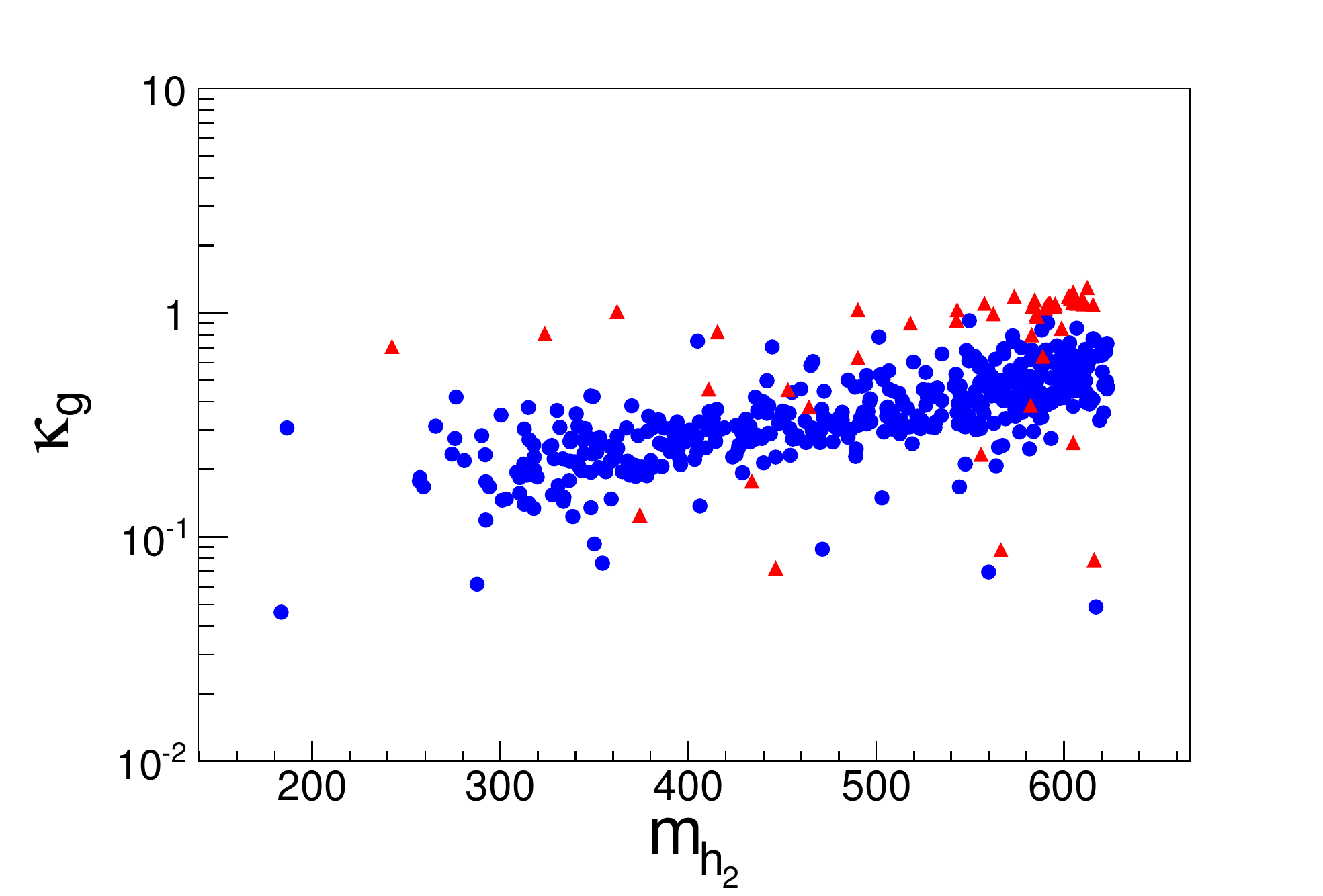}
\caption{The effective couplings of $h_1$ (left) and $h_2$ (right) to gluons $\kappa_g$ as a function of mass, where $\kappa_g$ is normalized to the gluon coupling of a SM Higgs boson at the same mass.}
\label{fig:kappag}
\end{centering}
\end{figure}

In Fig.~\ref{fig:kappag} we show the effective coupling to gluons for $h_1$ and $h_2$ versus their respective masses, normalized to the  gluon coupling of an SM-like Higgs at the same mass. 
We see that the couplings could be typically the order of the SM value or slightly larger for $h_1$ less than $90$ GeV. For $h_1$ above $90$ GeV the maximum coupling is decreasing from roughly $\sim 0.7$ to $\sim 0.3$ times the SM-like value as $h_1$ becomes heavier. This bound comes directly from existing searches above $M_Z$.
Further searches for resonances decaying to taus at the LHC will thus directly impact most of the model space. 
A preliminary update from CMS on such searches, which we have included, already significantly impacts our results \cite{Khachatryan:2014wca,CMStauupdate}. 
Given that searches for multiple tau final states are complicated, one can also consider final states involving a muon pair since the light neutral state will have a small but non-vanishing branching fraction $BR(\mu\mu) \sim 0.1\%$. For very light scalars this fraction can become much higher. 
One can also search for the heavy partner $h_2$, decaying either to $\tau\tau$, to $Z \tau \tau$, or to $4\tau$. As shown in the right panel of Fig.~\ref{fig:kappag}, $h_2$ may have gluon couplings between $\sim 0.01$ and $\sim 1$ compared to the SM-like case. The updated search for $h_2 \to h_1 Z \to \tau\tau ll$ already limits our surviving points and future results will directly impact the model \cite{HZtauupdate}.

For many of our viable points, the leading production mechanisms for new states at hadron colliders are associated with the heavy quarks
\be 
gg \to H/A\quad {\rm and}\quad b\bar b \to H/ A,\ \ gb \to t H^\pm.
\ee 
The gluon-fusion channel will continue to be a leading search mode at the LHC. However, given the wide range of the possible $\kappa_g$ values as seen in Fig.~\ref{fig:kappag}, more quantitative evaluation would be needed to draw a conclusion for the observability of those states at the LHC. 
The $b$-associated channel can be mildly enhanced when $|A_d| > 1$, 
but the $t$-associated channel will be suppressed due to $|A_u| < 1$. 

A second production mechanism is through the electroweak pair production channels
\be 
q\bar q' \to W^\pm \to H^\pm A,\ H^\pm H,\ H^\pm h,\ \quad {\rm and}\quad q\bar q \to Z \to AH,\ Ah.
\ee 
The charged current channel $W^\pm \to H^\pm A$ is via the pure SU(2) gauge interaction independent of other model parameters, and the process $Z \to AH$ is also close to the full EW strength due to the small $h-H$ mixing. By the same token, production of $h$ via these modes is quite small. These channels have the advantage of being present even if the quark couplings are small. 
It is conceivable to search for $ Z^* \to AH \to 4 \tau / 4\tau Z / 6\tau$, and similarly for $W^{+*} \to H^+ h_1 \to 3 \tau/4\tau W^+$ could be pursued. $W^{+*} \to H^+ h_2$ is also possible, although typically it will decay to a more complicated final state. Estimates of the sensitivity of such searches in the Type X model were recently published in Ref.~\cite{Chun:2015hsa}. The $Z^*$ channel can also be exploited at a lepton collider, particularly at the ILC which, with an energy of  $500$ GeV or $1$ TeV could kinematically access a large part of the interesting parameter space. 


\section{Summary and Conclusions}
\label{sec:conclude}

In this work we have explored the viable parameter space of the CP-conserving A2HDM, outlined in Sec.~\ref{sec:review},  which can account for the observed value of $\Delta a_\mu$. 
As studied in great detail in Sec.~\ref{sec:constraints}, the model is significantly constrained by experimental measurements of the $T$-parameter, by flavor changing neutral decays $B_s \to \mu^+\mu^-$ and $b \to s \gamma$, by precision measurement of the $Z\tau\bar{\tau}$ coupling and $\tau$ decays,  by the measured production and decay channels for the SM-like Higgs boson, and by LHC searches for exotic Higgs decays. 
We find in Sec.~\ref{sec:results} that it is possible to satisfy all our requirements in certain parameter regions with characteristic features of the new Higgs bosons.
Our results can be summarized as follows:
\begin{itemize}
\item There are three distinctive mass regions that are viable as given in Eq.~(\ref{eq:regions}). These are distinguished by the mass values of $H$ and $A$ and by their dominant contributions to $\gtwo$, as illustrated in Fig.~\ref{fig:diagrams}.
\item The Yukawa couplings of the new scalars to top quarks can play a leading role in generating the necessary  enhancement. This in turn allows for a wider range of masses compared to the Type X model, including  heavy pseudo-scalars, as well as very light scalars, with $m_H \lesssim 10$ GeV or $m_A^{} \sim 10$ GeV.
\item The new Higgs states all have large couplings to $\tau$'s due to the necessarily large value of $A_l$. 
This can have observable effects in precision measurements such as $Z \to \tau \overline{\tau}$, and universality in (semi-)leptonic decays, and it dominates the phenomenology of collider searches (Sec.~\ref{sec:SM}). 
\item Light new states ($h_1=A, H$) decay to tau pairs.  
Heavier states  
will typically decay to either a pair $h_1h_1$ or to $h_1 Z$, although a significant fraction may decay directly to a pair of taus or to tops. 
Thus, further searches for exotic Higgs decaying to taus or to taus plus $Z$ in the final state can potentially discover the new particles of this model (Sec.~\ref{sec:search}).
\item The leading production at the LHC may still be associated with the heavy quarks, such as $gg\to H/A,\ b\bar b\to H/A,\ gb \to tH^\pm$. New particles can also be pair produced through $Z^* \to AH$ and $W^* \to H^{+} (A/H)$ mainly via gauge interactions. They can be competitive in the search when $A/H$ is light  (Sec.~\ref{sec:search}).
\item The obtained solutions above may lead to significant modifications to the SM measurements, such as the deviation of the SM-like Higgs boson couplings, and
a new decay channel  to 4 taus (Sec.~\ref{sec:SMi}).
\item Current LHC searches have already begun to impact the allowed parameter space and the model is sensitive to future searches.
 
\end{itemize}

Our work demonstrates the viability of a much larger class of theory related to the 2HDM, which can explain $\gtwo$, with the Type X model as a special case within this framework.
The model presents interesting coupling patterns, new production channels for both the SM-like Higgs boson and the new states at hadron colliders, and a richer phenomenology at low energies. 

\appendix
\section{Calculation of Muon $g-2$}

Here we present formulae for the calculation of dominant terms in the new physics contributions to $g-2$.
One loop contributions to $\Delta a_\mu$ are given by the following expressions \cite{Dedes:2001nx,g-22hdm}:
\be
(\Delta a_\mu^1)^{\mbox{neutral}} &=& \sum_i \frac{ m_\mu^2}{8 \pi^2 v ^2 m_{h_i}^2}\left(  (\mbox{Re}(y_{L}^{h_i}))^2 F_H \left(\frac{m_\mu^2}{m_{h_i}^2}\right) + (\mbox{Im}(y_{L}^{h_i}))^2 F_A \left(\frac{m_\mu^2}{m_{h_i}^2}\right)\right)  \nonumber \\
(\Delta a_\mu^1)^{\mbox{charged}} &=& \frac{ m_\mu^2}{8 \pi^2 v^2 m_{H^+}^2} |A_l|^2 F_{H^+}\left(\frac{m_\mu^2}{m_{H^+}^2}\right)
\ee
where $i$ runs over the neutral Higgs eigenstates. The relevant functions are:
\be
F_H(z) = \int_0^1 dx \frac{x^2(2-x)}{zx^2-x+1}, \ \ 
F_A(z) = \int_0^1 dx \frac{-x^3}{zx^2-x+1}, \ \ 
F_{H^+}(z) = \int_0^1 dx \frac{-x^2(1-x)}{zx^2_(1-z)x}.  
\nonumber
\ee

The coupling of a neutral Higgs to fermions of a given type $f=t,~b,~\tau$ is determined by 
\be
y_f^{h_i}=R_{i1}+(R_{i2}\pm i R_{i3})(A_{d,l (u)}). 
\ee
 with the negative sign applying for up-type quarks. 

The two loop contributions to $\Delta a_\mu$ from Barr-Zee diagrams with a photon and a neutral Higgs are given by \cite{Cheung:2003pw}
\be
(\Delta a_\mu^2) = \sum_{f,i} \frac{N_c^f \alpha_{em} m_\mu^2 Q_f^2}{4 \pi^3 v^2} \left( \mbox{Im}(y_L^{h_i})\mbox{Im}(y_f^{h_i}) G\left(\frac{m_f^2}{m_{h_i}^2}\right) -\mbox{Re} (y_L^{h_i}) \mbox{Re} (y_f^{h_i}) F\left(\frac{m_f^2}{m_{h_i}^2}\right) \right),~~
\ee
where $N_c^f$ is the number of colors and $Q_f$ the charge of the fermion 
$f$ in the loop. 
The functions $G(z)$ and $F(z)$ are defined as:
\be
G(z) &=\frac{z}{2} \int_0^1 dx \frac{1}{x(1-x)-z} \ln \frac{x(1-x)}{z}, \ \ 
F(z) &=\frac{z}{2} \int_0^1 dx \frac{1-2x(1-x)}{x(1-x)-z} \ln \frac{x(1-x)}{z}.
\ee

Additional charged Higgs contributions are detailed in Ref.~\cite{Ilisie:2015tra}.

\acknowledgments

We would like to thank Eung Jin Chun and Tomohiro Abe for discussions. 
This work was supported in part by the U.S.~Department of Energy under grant No. DE-FG02-95ER40896, in part by the PITT PACC, and in part by the National Research Foundation of
Korea (NRF) grant funded by the Korea government of the Ministry of Education, Science
and Technology (MEST) (No. 2014R1A1A2057665).

\end{document}